\documentclass[aps,reprint,colorlinks=true, a4paper=true, pdfstartview=FitV,linkcolor=blue, citecolor=blue, urlcolor=blue,]{revtex4-1}

\usepackage[T1]{fontenc}		
\usepackage[english]{babel}		
\usepackage[utf8]{inputenc}	
\usepackage{times}
\usepackage{tikz}
\usepackage{tensor}
\usetikzlibrary{knots}

\usepackage{amsmath}			
\usepackage{amssymb}			
\usepackage{bbm}				

\usepackage{hyperref}
\usepackage{pbox}

\usepackage{graphicx}			
\usepackage{graphics}			
\DeclareGraphicsExtensions{.pdf}
\usepackage{color}		

\newcommand{\bea}{\begin{eqnarray}}
\newcommand{\eea}{\end{eqnarray}}

\DeclareMathOperator{\R}{\mathbb{R}}

\begin{document}

\title{$\mathcal{PT}$ symmetry-protected exceptional cones and analogue Hawking radiation}

\author{Marcus St{\aa}lhammar$^{1}$}
\author{Jorge Larana-Aragon$^1$}
\author{Lukas R\o dland$^1$}
\author{Flore~K. Kunst$^{2,3}$}

\affiliation{$^1$Department of Physics, Stockholm University, AlbaNova University Center, 106 91 Stockholm, Sweden}
\affiliation{$^2$Max Planck Institute of Quantum Optics, Hans-Kopfermann-Stra\ss e 1, 85748 Garching, Germany}
\affiliation{$^3$Max Planck Institute for the Science of Light, Staudtstra\ss e 2, 91058 Erlangen, Germany}

\begin{abstract}

Non-Hermitian Hamiltonians, which effectively describe dissipative systems, and analogue gravity models, which simulate properties of gravitational objects, comprise seemingly different areas of current research. Here, we investigate the interplay between the two by relating parity-time-symmetric dissipative Weyl-type Hamiltonians to analogue Schwarzschild black holes emitting Hawking radiation. We show that the exceptional points of these Hamiltonians form tilted cones mimicking the behavior of the light cone of a radially infalling observer approaching a black hole horizon. We further investigate the presence of tunneling processes, reminiscent of those happening in black holes, in a concrete example model. We interpret the non-trivial result as the purely thermal contribution to analogue Hawking radiation in a Schwarzschild black hole. Assuming that our particular Hamiltonian models a photonic crystal, we discuss the concrete nature of the analogue Hawking radiation in this particular setup.

\end{abstract}

\maketitle

\section{Introduction} 

Since the experimental discovery of the quantum Hall effect in 1980 \cite{QHE}, topology has played a prominent role in condensed matter physics. Gapped topological insulators \cite{hasankane,qizhang} and gapless semimetals, such as graphene \cite{goerbig} and Weyl semimetals (WSMs) \cite{weylreview}, have Bloch bands that are characterized by distinct topological invariants, which play a crucial role in establishing a bulk-boundary correspondence \cite{jantopreview}. Notably, the elusive Weyl fermions, despite evading discovery for nearly a century after being theoretically predicted in the context of particle  physics \cite{W1929}, are realized as quasi-particles near the point-like band crossings in WSMs \cite{XBAN2015,LWFM2015,LWYRFJS2015}. 
This has led to further observations of theoretical concepts known from high-energy physics, such as the chiral anomaly, which has been observed in experiments \cite{Chiral1,Chiral2,Chiral3,Chiral4}.

The general idea of establishing analogies between different areas of physics is especially relevant in the context of black holes. Even though, the study of black holes has provided key theoretical insights into the fundamental nature of spacetime, direct experimental input is difficult to obtain. However, through investigations in analogue models, black hole physics can be studied in laboratory setups \cite{Unruh,Barcelo}.
In this context, WSMs have received great attention due to the behavior of their cone-like dispersion, called the Weyl cone. In particular, the tilting of Weyl cones resembles the tilting of spacetime light cones of observers in the presence of a black hole \cite{beenakker}. This resemblance has been formalized via the construction of a relation between the transition from type-I to type-II WSMs, where the Weyl cones are overtilted [cf. Fig.~\ref{fig:Wcones}(d)], and the tilting of an observer’s light cone when crossing the black hole horizon in the infalling coordinate frame \cite{Volovik2016}. With this as a starting point, recent studies \cite{volovik,zobkovblackholes,Zubkovold,Liu2018,Yaron2020,Nissinen} have focused on investigating potential analogues of Hawking radiation in WSMs: In 1975, Hawking postulated that despite the classical intuition, black holes emit thermal radiation at a characteristic temperature, the Hawking temperature, when taking quantum effects into account, and thus evaporate \cite{Hawking1975}. While some of these works predict the existence of analogue Hawking radiation in WSMs \cite{volovik, Liu2018,Zubkovold}, others find no radiative processes \cite{Yaron2020}.
Indeed, in Ref.~\onlinecite{Yaron2020} the absence of analogue Hawking radiation is attributed to the stability of the Hermitian Weyl model thus motivating studies of black hole analogue models beyond these systems.

Inspired by the fact that the emission of thermal radiation can be associated with the loss of energy from a system to its environment, we here instead consider a dissipative Weyl-type model with parity-time ($\mathcal{PT}$) symmetry described by a non-Hermitian (NH) Hamiltonian. NH Hamiltonians serve as effective descriptions of systems subject to gain and loss \cite{NHreview}, and find applications both in classical \cite{Lin2011, Regensburger2012, Peng2014, Feng2014, Hodaei2014, Peng2016, Hodaei2017, Chen2017, Ozdemir2019, Fleury2015, Helbig2020, Ghatak2020} and quantum setups \cite{Kreibich2014, koziifu, Yoshida2018, Yoshida2020, Bergholtz2019}.
Naturally, these Hamiltonians are fundamentally different from their Hermitian counterparts, the most prominent distinction being that the  eigenvalues are generally complex with the imaginary part associated with finite lifetimes and the sets of left and right eigenvectors may be unequal. Furthermore, the nodal structures of NH systems consist of exceptional points (EPs) at which not only the eigenvalues, but also the eigenvectors coalesce \cite{brody14}. The sets of EPs exhibit fascinating topological properties \cite{BerryDeg,carlstroembergholtz,molina,disorderlinesribbons,koziifu,ourknots,Ronnyknot1,ourknots2,symprotnod}, which can be investigated in experiments at remarkable precision \cite{Lin2011, Regensburger2012, Peng2014, Feng2014, Hodaei2014, Peng2016, Hodaei2017, Chen2017, expknots}. In particular, NH systems find high experimental relevance in optics, where they effectively describe photonic systems subject to $\mathcal{PT}$ symmetry \cite{Ozdemir2019, topphot,speclat,Ozawa2019}. $\mathcal{PT}$-symmetric systems host distinct regions with real and complex eigenvalues, which are separated by EPs, where $\mathcal{PT}$ symmetry is spontaneously broken \cite{Bender1998}. Consequently, these regions are called the $\mathcal{PT}$-unbroken and the $\mathcal{PT}$-broken phase, respectively.

Our dissipative Weyl-type model features cones in the spectrum consisting of EPs, which we call \emph{exceptional cones} (ECs), whose intersection point, i.e., where the vertices of two cones touch, corresponds to an ordinary node. For this setup, we investigate whether a possible relation exists between ECs and spacetime light cones near a black hole. In particular, we show that ECs of order two, where the order refers to the number of eigenvectors that coalesce, appear generically in a family of two-band, three-dimensional (3D) $\mathcal{PT}$-symmetric models. These ECs look similar to Weyl cones but possess additional features. For example, the EC separates a region, where the eigenvalues are real, from a region, where the eigenvalues are imaginary thus mimicking the space-like and time-like regions, respectively, that are separated by an observer's spacetime light cone.

By mapping the EC onto a 2D Weyl cone, we make a formal connection between the EC and an observer's light cone in the vicinity of a stationary (3+1)D Schwarzschild black hole. We then investigate the existence of analogue Hawking radiation by calculating the semiclassical quantum tunneling rate. Indeed, following previous approaches where Hawking radiation is regarded as a quantum tunneling process \cite{Volovik1999, PW2000}, we calculate the leading order contributions to analogue Hawking radiation and find them to be nonzero.  To address the experimental relevance of our findings, we show that our Hamiltonian model can be thought of as an effective description derived from the Lindblad master equation describing a photonic crystal. We then discuss the explicit nature of the analogue Hawking radiation, including its source and how the pair-production mechanism works, suggesting that these systems constitute promising candidates in which the analogy can be tested.

This paper is setup as follows: Due to the interdisciplinary character of this article, we start with a brief discussion on Hawking radiation, WSMs and NH systems in Sec.~\ref{sect:preliminaries}, which are concepts necessary for understanding the present work. This is followed by the introduction of our model and the establishment of a relation to the spacetime light cone in Sec.~\ref{sec:cones}. In Sec.~\ref{sec:partemission}, we apply the quantum tunneling method in order to compute the analogue semiclassical emission rate, and we discuss possible experimental realizations of our model in Sec.~\ref{sec:experiment}. We conclude in Sec.~\ref{sec:conclusions}.

Throughout this article, we use Planck units, i.e., $\hbar=c=k_{\text{B}}=G_{\text{N}}=1$, such that all physical quantities become dimensionless.

\section{Preliminaries} \label{sect:preliminaries}
Here we introduce several concepts important for our work. In Sec.~\ref{sec:BHrad}, we briefly discuss black holes and Hawking radiation serving as complementary motivation for constructing analogue gravity models. In Sec.~\ref{sec:Hmodels}, we present the relation between \emph{Hermitian} two-band models and curved spacetimes, especially highlighting WSMs as gravitational analogue models. Finally, we discuss NH two-band models and the role played by $\mathcal{PT}$ symmetry in Sec~\ref{sec:NHmodels}. The notation and conventions introduced here are used throughout the paper.

\subsection{Hawking radiation and black holes} \label{sec:BHrad}
Black holes are regions in spacetime composed of a singularity, which is a point where the curvature of spacetime becomes infinite, and a boundary surface, known as the event horizon. This is clearly illustrated in the case of the most general static, spherically symmetric solution to Einstein's equations in vacuum, i.e., the asymptotically flat (3+1)D \emph{Schwarzschild black hole} \cite{Schwarzschild}. The most common way of writing the metric for this type of black hole is in terms of Schwarzschild coordinates, but these cover only the region outside the black hole horizon and lack a description of the interior. For later purposes, which will become clear in Sec.~\ref{sec:partemission}, we here instead choose to represent the metric with a set of coordinates valid both in the inner and outer region of the black hole as well as at its horizon, such as the Painlev\'e-Gullstrand coordinates \cite{Painleve,Lemaitre}. In this coordinate frame, the line element reads
\begin{equation} \label{eq:painleve}
    ds^2 = -\left(1-\frac{2M}{r}\right)dT^2+2\sqrt{\frac{2M}{r}}drdT + dr^2+r^2d\Omega^2,
\end{equation}
where $M$ is the mass of the black hole, $r$ is the Painlev\'e radial coordinate, $T$ is the Painlev\'e time and $d\Omega$ is the differential element of solid angle of the 2-sphere. Evidently, Eq.~\eqref{eq:painleve} has a mathematical singularity at $r=0$ corresponding to the physical singularity of the black hole. The event horizon, on the other hand, is located at $r=2M$ where the time component of Eq.~\eqref{eq:painleve} vanishes. In Schwarzschild coordinates, this point corresponds to a coordinate singularity, while in the chosen coordinate frame the metric is regular at that point. The intuition for the Painlev\'e-Gullstrand coordinates naturally arises from considering radial null geodesics of infalling observers into the black hole \cite{Vanzo2011}. These can be obtained from the metric above and read
\begin{equation}
\dot{r}_{\pm}=\pm1-\sqrt{\frac{2M}{r}},
\end{equation}
where $\dot{r}:=dr/dT$. By studying the asymptotic behavior of $\dot{r}_{\pm}$, we see that $\dot{r}_{+}$ and $\dot{r}_{-}$ correspond to outgoing and ingoing solutions as $ \lim_{r\to +\infty}\dot{r}_+>0$ and $\lim_{r\to +\infty}\dot{r}_-<0$, respectively. Indeed, these coordinates describe infalling observers whose light cones tilt as they approach the black hole horizon. The light cone reaches a critical tilting value at the horizon and it overtilts in the interior, where the roles of time and space are interchanged. This occurs generically for infalling observers approaching a black hole horizon, and can also be seen in, e.g., Eddington-Finkelstein coordinates \cite{carroll}.

Classical physics dictates that nothing, not even light, can escape from a black hole once it has crossed the event horizon. In sharp contrast, when taking quantum fluctuations into account, black holes are predicted to emit thermal radiation at a characteristic temperature, called Hawking radiation and the Hawking temperature, respectively, resulting in their evaporation. Let us illustrate this by considering the case of a Schwarzschild black hole. The phenomenon of Hawking radiation can be seen explicitly from the expression for the spectrum of Hawking quanta \cite{Hawking1975}, which resembles a Planckian spectrum
\begin{equation}\label{eq:blackbodyspectrum}
    N(\omega) =\frac{\gamma_{\omega}}{e^{\beta\omega}-1},
\end{equation}
where $N$ represents the total number of emitted quanta of energy $\omega$, $\beta=1/T_{\text{BH}}$ is the inverse Hawking temperature for a Schwarzschild black hole
\begin{equation}\label{eq:Hawkingtemp}
    T_{\text{BH}}=\frac{1}{8\pi M},
\end{equation}
and $\gamma_{\omega}$ is a classical absorption coefficient, known as the greybody factor, which makes the spectrum deviate slightly from the one characteristic of a blackbody. For sufficiently massive black holes the thermal approximation is justified since the greybody factor can be neglected \cite{JennieTraschen}, i.e., $\gamma_{\omega}\rightarrow1$ when $\left|\omega\right|M\sim\left|\omega\right|/T_{\text{BH}} \gg 1$, and Eq.~\eqref{eq:blackbodyspectrum} can be approximated by the Boltzmann factor $e^{-\beta\omega}$. Since the Hawking temperature $T_{\text{BH}}$ is inversely proportional to the black hole mass $M$, the evaporation process is an accelerated process in the sense that the black hole becomes hotter as it emits (almost purely) thermal radiation. Following this reasoning, it is possible to consider black holes as thermal systems with associated thermodynamic variables. The quantity of interest when it comes to Hawking radiation is the semiclassical emission rate \cite{KrausKeski}
\begin{equation}\label{eq:evaporationrate1}
\Gamma(\omega)\propto N(\omega)  \sim e^{-\beta\omega},
\end{equation}
which quantifies the rate of emission of thermal radiation away from the black hole.

The phenomenon of Hawking radiation is very elusive in the sense that its direct detection in astrophysical black holes is highly masked by the cosmic microwave background, and quantities such as the emission rate can thus not be measured.
As such, Hawking radiation has never been directly observed.
This motivates the field of analogue gravity in the search for analogue black hole models with the aim to gain insight into the nature of Hawking radiation.

In this work, we follow the interpretation of Parikh and Wilczek in Ref.~\onlinecite{PW2000}, who treat Hawking radiation as a quantum tunneling process.

\subsection{Weyl semimetals and analogue spacetime light cones} \label{sec:Hmodels}
\begin{figure*}[hbt!]
\centering
\includegraphics[width=\textwidth]{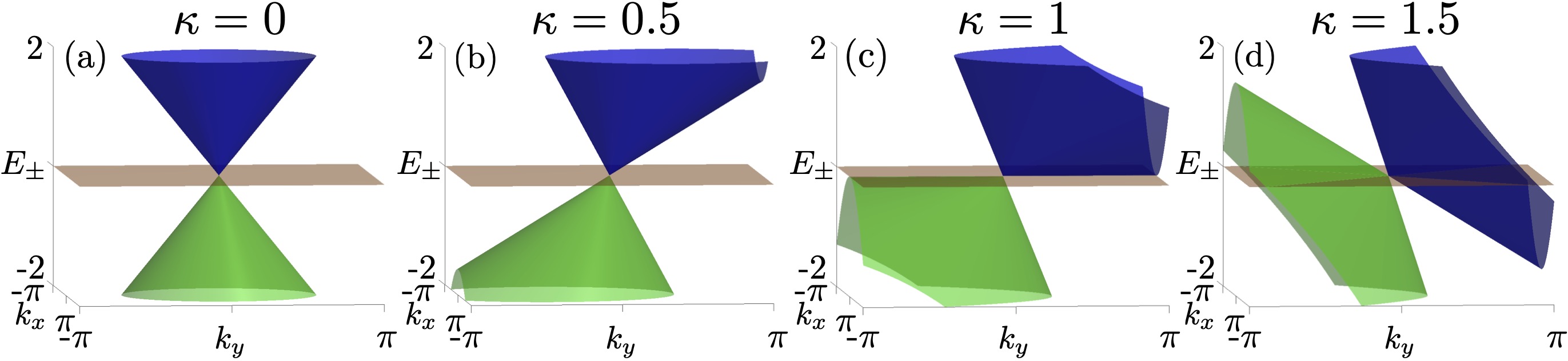}
\caption{Illustration of the eigenvalues corresponding to Eq.~\eqref{eq:WHam} for $\boldsymbol{\kappa}=\left(\kappa,0,0\right)$ in the plane $k_z=0$ for different values of $\kappa$: (a) $\kappa = 0$, (b) $0.5$, (c) $1$, and (d) $1.5$. When $\kappa> 1$, the cone is overtilted in analogy to the light cone of an observer that has crossed the black hole horizon. Thus, the two phases, type I ($\kappa<1$) (a,b) and type II ($\kappa>1$) (d), intuitively resemble the exterior and the interior of a black hole, respectively.}\label{fig:Wcones}
\end{figure*}
We continue by describing the connection between Weyl cones and spacetime light cones in Hermitian systems. A non-interacting Hermitian two-band model in its most general form is given by the following Bloch Hamiltonian
\begin{equation}\label{eq:eq1}
    \mathcal{H}(\mathbf{k}) = d_0(\mathbf{k}) \sigma^0 + \mathbf{d}(\mathbf{k})\cdot \boldsymbol{\sigma},
\end{equation}
where $\mathbf{k}$ is the lattice momentum with the appropriate dimensions, $\sigma^0$ is the $2\times 2$ identity matrix, $\boldsymbol{\sigma} = \left(\sigma^x,\sigma^y,\sigma^z\right)$ is the vector of Pauli matrices, $\mathbf{d}(\mathbf{k}) = \left[d_x(\mathbf{k}),d_y(\mathbf{k}),d_z(\mathbf{k})\right]$, and $d_{\mu}(\mathbf{k})$ are continuously differentiable real-valued functions of $\mathbf{k}$ for $\mu = 0,x,y,z$. From here on, we use the notation $d_{\mu}:=d_{\mu}(\mathbf{k})$ for simplicity, while restoring the $\mathbf{k}$-dependence when appropriate. The corresponding energy eigenvalues are
\begin{equation}
    E_{\pm} = d_0 \pm \sqrt{d_x^2+d_y^2+d_z^2}.
\end{equation}
Nodal points or higher-dimensional degenerate structures are given by the intersections of the eigenvalues, which arise from solving
\begin{equation} \label{eq:nodalpoints}
    \sqrt{d_x^2+d_y^2+d_z^2}=0.
\end{equation}
Solutions to this equation are of codimension 3 meaning that nodal points occur generically in three dimensions or higher. Well-studied examples, where these nodal points appear, include WSMs, whose dispersions form cones close to the nodal points, referred to as Weyl cones. This intriguing feature, together with the relativistic aspect of the Weyl nodes, led to WSMs being proposed as potential analogue gravity systems by relating Weyl cones to spacetime light cones \cite{Volovik2016,volovik,zobkovblackholes,Zubkovold,Liu2018,Yaron2020, Nissinen}. The starting point of this analogy is the Weyl Hamiltonian
\begin{equation} \label{eq:WHam}
    \mathcal{H} = -\boldsymbol{\kappa} \cdot \mathbf{k} \sigma^0+\mathbf{k}\cdot\boldsymbol{\sigma},
\end{equation}
where $\boldsymbol{\kappa} \in \R^3$ is a tilting parameter and $\mathbf{k}:=\left(k_x,k_y,k_z\right)$. The corresponding energy eigenvalues read
\begin{equation} \label{eq:genWcone}
    E_{\pm} = -\boldsymbol{\kappa} \cdot \mathbf{k}\pm \sqrt{k_x^2+k_y^2+k_z^2},
\end{equation}
and constitute a cone in energy-momentum space tilted by $\boldsymbol{\kappa}$, which is displayed in Fig.~\ref{fig:Wcones}. $|\boldsymbol{\kappa}|<1$ and $|\boldsymbol{\kappa}|>1$ correspond to type I and type II WSMs, respectively. The transition between these two different types, i.e., when the cone overtilts, is intuitively similar to the tilting of the light cone when the corresponding radially infalling observer crosses the horizon of a Schwarzschild black hole.

This intuition is made formal by relating the Hamiltonian~\eqref{eq:WHam} to the metric of a curved spacetime. We note that such a relation exists for any Dirac-like operator, and we refer to Refs.~\onlinecite{Volovik2014,Horova2005} for a complete and detailed treatment. The key ingredient is the so-called \emph{vielbeins} $e\indices{^{\mu}_{\alpha}}$ and their inverses $e\indices{_{\mu} ^{\alpha}}$, which in general relativity are used to define local patches of orthonormal frames in spacetime \cite{Ortin,carroll}. They relate to the (inverse) spacetime metric as $g^{\mu\nu}= e\indices{^{\mu}_{\alpha}}e\indices{^{\nu}_{\beta}}\eta^{\alpha\beta}$, with $\eta^{\alpha\beta}=\text{diag}\left(-1,1,1,1\right)$. Following Ref.~\onlinecite{Volovik2014}, the Hamiltonian in Eq.~\eqref{eq:WHam} can be put on the form
\begin{equation} \label{eq:vielbeinHerm}
    \mathcal{H} = e\indices{^i_a}k_i\sigma^a+e\indices{^i_0}k_i\sigma^0,
\end{equation}
with $i,a=x,y,z$, $e\indices{^i_a}=\delta\indices{^i_a}$, and $e\indices{^i_0}=-\kappa^i$. Setting $e\indices{^0_0}=1$ \cite{Volovik2014,Volovik2016}, we find that the line element of the corresponding analogue spacetime metric, $ds^2 = g_{\mu \nu} dx^\mu dx^\nu$, reads
\begin{equation} \label{eq:analmetricherm}
    ds^2 = -\left(1-\left|\boldsymbol{\kappa}\right|^2\right) dt^2+\left|d\mathbf{x}\right|^2+2\boldsymbol{\kappa} \cdot d\mathbf{x}\,dt,
\end{equation}
where $d\mathbf{x} = \left(dx,dy,dz\right)$ and $|d\mathbf{x}|^2 = dx^2+dy^2+dz^2$. Eq.~\eqref{eq:analmetricherm} describes an event horizon at $|\kappa|=1$, which coincides exactly with when the Weyl cone overtilts, cf. Fig.~\ref{fig:Wcones}.
Next, we present a short description of NH systems and the role played by $\mathcal{PT}$ symmetry.

\subsection{NH systems and $\mathcal{PT}$ symmetry} \label{sec:NHmodels}
The mathematical description of systems effectively modelled by NH Hamiltonians is a straightforward generalization of the reasoning presented in Sec.~\ref{sec:Hmodels}. We again consider the Hamiltonian in Eq.~\eqref{eq:eq1}, where $d_{\mu}$ now denote \emph{complex-valued} continuously differentiable functions of the lattice momentum $\mathbf{k}$. Decomposing $\mathbf{d} = \mathbf{d}_{\text{R}}+i\mathbf{d}_{\text{I}}$ with $\mathbf{d}_{\text{R}}$ and $\mathbf{d}_{\text{I}}$ denoting the real and imaginary parts of $\mathbf{d}$, respectively, the corresponding complex eigenvalues are
\begin{equation}
    E_{\pm} = d_0 \pm \sqrt{d_{\text{R}}^2-d_{\text{I}}^2+2i\mathbf{d}_{\text{R}}\cdot \mathbf{d}_{\text{I}}}.
\end{equation}
Hence, the nodal structure is given by solutions to the following system of equations,
\begin{equation} \label{eq:expoint}
    d_{\text{R}}^2-d_{\text{I}}^2=0, \qquad \mathbf{d}_{\text{R}}\cdot\mathbf{d}_{\text{I}}=0.
\end{equation}
In contrast to the eigenvalue degeneracies in Hermitian systems, the nodal points in NH systems are \emph{exceptional points}. The exceptional nodal structure is determined by the simultaneous solution of two equations meaning that it has codimension 2. Stable EPs thus occur already in 2D systems, while in 3D solutions to Eq.~\eqref{eq:expoint} attain the form of closed lines \cite{NHreview}. The EPs furthermore constitute the boundary of topologically protected bulk states, called Fermi states (FS) and i-Fermi states (iFS). These correspond to the regions in momentum space defined by
\begin{align}
    \text{Re}\left(E_+\right)=\text{Re}\left(E_-\right) \Rightarrow \mathbf{d}_{\text{R}}\cdot\mathbf{d}_{\text{I}} &= 0, \quad d_{\text{R}}^2-d_{\text{I}}^2 \leq 0,
    \\
    \text{Im}\left(E_+\right)= \text{Im}\left(E_-\right) \Rightarrow \mathbf{d}_{\text{R}}\cdot\mathbf{d}_{\text{I}} &= 0, \quad d_{\text{R}}^2-d_{\text{I}}^2 \geq 0,
\end{align}
respectively. The presence of these (i)FSs in the bulk of the spectrum is a characteristic signature of dissipation, and has been experimentally observed in photonic lattices \cite{bfermiarcs}.

Imposing $\mathcal{PT}$ symmetry constrains the Hamiltonian as follows. Following Refs.~\onlinecite{BLC1,BLC2}, we represent $\mathcal{P}$ symmetry by a $\mathcal{P}$-matrix and $\mathcal{T}$ symmetry by a $\mathcal{K}$-matrix, where $\mathcal{KK}^*=\pm \mathbb{I}$, $\mathcal{P}^2=\mathbb{I}$ and $\mathcal{KP}^*=\pm\mathcal{PK}$. We focus on the symmetry class where $\mathcal{KK}^*=\mathbb{I}$ and $\mathcal{KP}^*=\mathcal{PK}$, and choose the following representation
\begin{equation} \label{eq:PTrep}
    \mathcal{P}= \sigma^y, \qquad \mathcal{K}=\sigma^z.
\end{equation}
Imposing this choice of $\mathcal{P}\mathcal{T}$ symmetry on the generic two-band Hamiltonian in Eq.~\eqref{eq:eq1} results in the following constraint
\begin{equation} \label{eq:PT-symmetry}
    \mathcal{H}_{\mathcal{P}\mathcal{T}} = \mathcal{P}\mathcal{K} \mathcal{H}_{\mathcal{P}\mathcal{T}}^*\left(\mathcal{P}\mathcal{K}\right)^{-1},
\end{equation}
such that $d_0,d_x,d_y\in \mathbb{R}$ and $d_z\in i\mathbb{R}$. This representation of $\mathcal{PT}$ symmetry is used throughout the present work. A $\mathcal{PT}$-symmetric two-band Hamiltonian is thus necessarily on the form
\begin{equation}
\label{eq:NH_PT_symm_ham}
    \mathcal{H}_{\mathcal{PT}} = d_0\sigma^0 + d_x\sigma^x+d_y\sigma^y+i\tilde{d}_z \sigma^z,
\end{equation}
with $d_0, d_x,d_y,\tilde{d}_z\in \R$ and $i\tilde{d}_z:=d_z$. The corresponding eigenvalues then read
\begin{equation}
\label{eq:NH_PT_symm_eigs}
    E_{\pm, \mathcal{PT}} = d_0 \pm \sqrt{d_x^2+d_y^2-\tilde{d}_z^2}.
\end{equation}
As such, the exceptional nodal structures are defined by solutions to only one equation,
\begin{equation} \label{eq:PTsymEP}
    d_x^2+d_y^2=\tilde{d}_z^2,
\end{equation}
and stable EPs protected by $\mathcal{PT}$ symmetry occur already in 1D systems, exceptional lines in 2D and exceptional surfaces in 3D \cite{symprotnod}. This increased dimension of nodal structures can also be found in Hermitian systems subject to certain symmetries \cite{herring,branchCutSemimetals,Nodallinksemimetals,explink,floquetlinks,nodalknotsemimetals}. The EPs in the $\mathcal{PT}$-symmetric systems separate the eigenvalue spectrum into two different phases: the $\mathcal{PT}$-unbroken phase for $\tilde{d}_z^2<d_x^2+d_y^2$, where the eigenvalues are purely real, and the $\mathcal{PT}$-broken phase with $\tilde{d}_z^2>d_x^2+d_y^2$, where the eigenvalues are complex and appear in complex conjugate pairs. Exactly on the EPs, $\mathcal{PT}$ symmetry is spontaneously broken.

\section{$\mathcal{PT}$ Symmetry-Protected Exceptional Cones and Analogue Light Cones } \label{sec:cones}
\begin{figure*}[t]
\centering
\includegraphics[width=\textwidth]{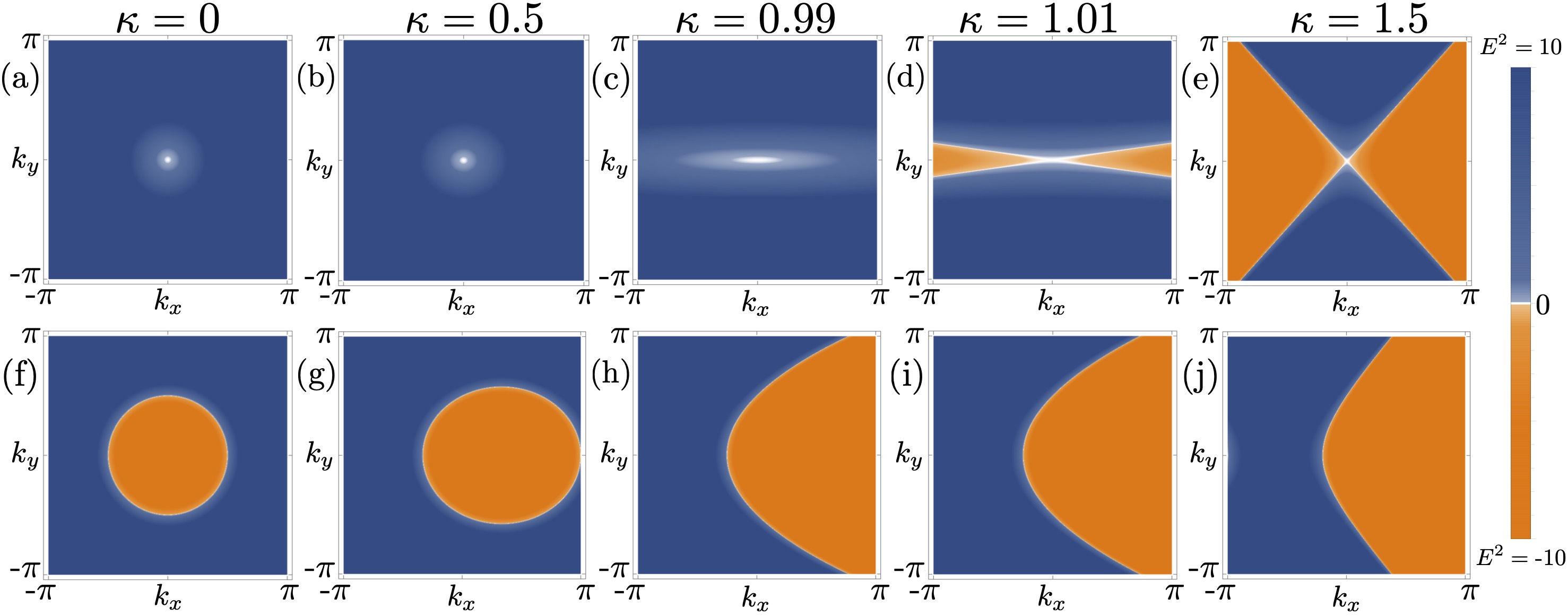}
\caption{(a)-(e) FS ($E^2<0$) and iFS ($E^2>0$), or equivalently, the $\mathcal{PT}$-broken and $\mathcal{PT}$-unbroken phases, respectively, of the model given by Eq.~\eqref{eq:NHham} at $k_z=0$ for different values of $\kappa$. For $\kappa<1$, the iFS constitute the whole Brillouin zone slice with a clear transition occurring at $\kappa=1$. This indicates the overtilting of the EC. In (f)-(j), the same is displayed at $k_z=\frac{\pi}{2}$.}\label{fig:FSurflin}
\end{figure*}

We now extend the analogy between spacetime light cones and WSMs (cf. Sec.~\ref{sec:Hmodels}) beyond the Hermitian case to the NH realm. In doing so, it is important to note that the relation between Dirac operators and spacetime metrics cannot be naively extended to NH operators. Additionally, the eigenvalues of NH operators are complex, such that there is no clear intuitive picture to follow as in the case of the (Hermitian) Weyl Hamiltonian. Interestingly, we show here that these challenges can be circumvented by studying the degeneracies in the eigenvalues of the NH $\mathcal{PT}$-symmetric Hamiltonian in Eq.~\eqref{eq:NH_PT_symm_ham}. This model hosts exceptional nodal structures determined by solving Eq.~\eqref{eq:PTsymEP}. If we now consider the $\mathcal{PT}$-symmetric NH Hamiltonian to be in three dimensions and at most \emph{linear} in momentum making the Hamiltonian Weyl-like, we find that (possibly tilted) cones appear, which are composed of EPs (we refer to Appendix~\ref{app:conerelation} for a general proof). These cones constitute stable exceptional structures in the sense that they are protected by the imposed $\mathcal{PT}$ symmetry. In particular, solving Eq.~\eqref{eq:PTsymEP} for $\tilde{d}_z$ gives
\begin{equation}
\label{eq:gen_sol_ECs}
    \tilde{d}_z=\pm \sqrt{d_x^2+d_y^2},
\end{equation}
which is on the same form as the eigenvalues of a \emph{Hermitian} Dirac operator. It further indicates that the EC could intuitively be interpreted as an analogue light cone.
In fact, this relation can be readily formalized at the level of equations, which we illustrate below for a concrete model with a generalization presented in Appendix~\ref{app:conerelation}.

Let us now turn to a specific example. Consider the NH $\mathcal{PT}$-symmetric two-band model
\begin{equation} \label{eq:NHham}
    \mathcal{H} = k_x\sigma^x+k_y\sigma^y + i\left(k_z+\kappa k_x\right)\sigma^z,
\end{equation}
with its corresponding eigenvalues
\begin{equation} \label{eq:evnh}
    E_{\pm}=\pm \sqrt{k_x^2+k_y^2-\left(k_z+\kappa k_x\right)^2}.
\end{equation}
Here, the term proportional to the identity matrix $d_0$ has been neglected since it does not affect the set of EPs. The EPs are explicitly given by
\begin{equation} \label{eq:exconekz}
    k_z = -\kappa k_x \pm \sqrt{k_x^2+k_y^2}.
\end{equation}
The latter constitutes a cone of EPs in momentum space tilted in the $k_x$-direction by the parameter $\kappa$. The corresponding squared eigenvalues are displayed in Fig.~\ref{fig:FSurflin}, highlighting the $\mathcal{PT}$-broken and $\mathcal{PT}$-unbroken phases. When $|\kappa|>1$, the corresponding (i)FSs become infinite [cf. Figs.~\ref{fig:FSurflin} (d),(e),(i),(j)], indicating that the EC is overtilted. Furthermore, similar to how an observer's spacetime light cone divides space-like and time-like regions, the EC separates regions of real and imaginary eigenvalues.

We note that $k_z$ in Eq.~\eqref{eq:exconekz} takes the same form as the eigenvalues of a Dirac-like operator, for instance on the form,
\begin{equation} \label{eq:Dirac op}
  \hat{k}_z=-\kappa k_x\sigma^0 + k_x\sigma^x + k_y\sigma^y. 
\end{equation}
Importantly, this equation and other Dirac-like operators obtained through this method are necessarily $\mathcal{PT}$-symmetric.
The connection between the EC and the eigenvalues of a Dirac-like operator, such as $\hat{k}_z$, allows us to relate the EC to a spacetime light cone. Recalling Sec.~\ref{sec:Hmodels},  $\hat{k}_z$ can be associated to a spacetime metric via the vielbein formalism,
\begin{equation}
    \hat{k}_z=e\indices{^i_a}k_i\sigma^a+e\indices{^i_0}k_i\sigma^0,
\end{equation}
with $i,a=x,y$, $e\indices{^i_a}=\delta\indices{^i_a}$, $e\indices{^x_0}=-\kappa$. From this and using $e\indices{^0_0}=1$, we find the following line element
\begin{equation} \label{eq:analmetricexam}
    ds^2 = -\left(1-\kappa^2\right)dt^2 + dx^2+dy^2+2\kappa dxdt,
\end{equation}
which is the analogue spacetime metric associated to the EC described by Eq.~\eqref{eq:exconekz}. Importantly, Eq.~\eqref{eq:analmetricexam} describes an event horizon at $|\kappa|=1$, which coincides with the overtilting of the EC. Since there are no relevant dynamics in the $y$ direction, this spurious degree of freedom can be eliminated from the analogue metric, and hence we set $dy=0$. This is a direct consequence of aligning the tilt of the EC in the $k_x$-direction. This metric has the same form as the one describing the spacetime of a radially infalling observer in the vicinity of a Schwarzschild black hole in Painlev\'e-Gullstrand coordinates. Indeed, we retrieve precisely the metric written in Eq.~\eqref{eq:painleve} when keeping the angular degrees of freedom constant, i.e., $d \Omega=0$, and identifying $x:=r$, $t:=T$ and $\kappa := +\sqrt{\frac{2M}{r}}$. We note that this relation between the EC and the spacetime light cone can be readily generalized to any 3D $\mathcal{PT}$-symmetric NH model with terms linear in momentum as presented in Appendix~\ref{app:conerelation}.

At this point, we need to comment on a caveat. To make the connection to the line element in Eq.~\eqref{eq:painleve}, we have promoted the tilting parameter to a function with explicit spatial dependence $\kappa=\kappa(r)$, and consequently, neither $\kappa(r)$ nor $k_r$ commute with the Hamiltonian. In principle this gives commutator contributions of the form $\left[k_r,\kappa(r)\right]\propto \frac{\partial \kappa(r)}{\partial r}$ when solving the eigenvalue equation of the Hamiltonian in Eq.~\eqref{eq:NHham}. In the following section, we provide a rigorous argument for why these contributions can be consistently neglected in the current study.

We observe that the expression for the Dirac-like operator $\hat{k}_z$ in Eq.~\eqref{eq:Dirac op} is identical to the Hamiltonian of a Hermitian 2D Weyl cone with a tilting term in the $k_x$-direction mediated by $\kappa$, and thus constitutes a 2D version of the Hamiltonian in Eq.~\eqref{eq:WHam}. Indeed, the line element in Eq.~\eqref{eq:analmetricexam} is identical to the line element in Eq.~\eqref{eq:analmetricherm} for ${\bf x} = (x ,y)$ and $\boldsymbol\kappa = \kappa \hat{x}$. This is not a surprising result, as we are relating one cone to another cone similar to what is done in Hermitian models. However, the conceptual difference between the EC and an ordinary Weyl cone leads to a very different interpretation of the physics they display.

We emphasize that there are additional symmetries that can be used to stabilize a cone of EPs. One such example is pseudo-Hermiticity, which at the level of exceptional eigenvalue degeneracies is equivalent to $\mathcal{PT}$ symmetry \cite{HOEP}. As previously mentioned, $\mathcal{PT}$ symmetry is of high experimental relevance in optics, and thus choosing $\mathcal{PT}$ symmetry instead of, e.g., pseudo-Hermiticity will be proven useful when we in Sec.~\ref{sec:experiment} discuss how our analogy could be experimentally tested.

\section{Spontaneous Particle Emission from Quantum Tunneling} \label{sec:partemission}

Having formalized the analogy between ECs and spacetime light cones in the previous section, we now turn to investigate physical properties of $\mathcal{H}$ in Eq.~\eqref{eq:NHham}. Recalling the motivation to extend the already existing analogies between Hermitian WSMs and light cones to include NH systems, we are primarily interested in lossy or dissipative features resembling analogue Hawking radiation.

To leading order in $M$ the contribution to Hawking radiation comes from light-like modes \cite{PageI,PageII}. For astrophysical black holes, for which $M$ is very large, these are dominating, while the contribution from massive modes are subleading \cite{massHR}. The former can be calculated using the quantum tunneling method near the horizon $r\sim 2M$. This treatment was first shown by Parikh and Wilczek, where the semiclassical emission rate is computed as a quantum tunneling rate \cite{PW2000}. Drawing intuition from that massless modes propagate along the light cone, and that the current analogy relates the light cone to the EC, we here apply the tunneling technique to the EC of the model in Eq.~\eqref{eq:NHham}. Thus, we calculate the semiclassical tunneling rate at the EC and interpret it as analogue Hawking radiation. The non-trivial result indicates that tunneling processes mimicking those sourcing the thermal contribution to Hawking radiation in the analogue black hole are present in the model.

\subsection{Commutator contributions near the horizon} \label{sec:commutator}

Before we apply the quantum tunneling method to our model, we first want to comment on the fact that the explicit spatial dependence of the tilting parameter $\kappa$ makes the commutators $\left[\mathcal{H},\kappa(r)\right]$ and $\left[\mathcal{H},k_r\right]$ non-vanishing. Consequently, when solving the corresponding eigenvalue equation additional contributions of the form $\left[k_r,\kappa(r)\right]$ will appear, which were neglected when deriving Eqs.~\eqref{eq:evnh} and \eqref{eq:exconekz}. The neglected terms evaluate to
\begin{equation} \label{eq:commutator}
    k_r\left[k_r,\kappa(r)\right] \propto k_r\frac{\partial\kappa(r)}{\partial r}\propto k_r\frac{M}{r^2\sqrt{\frac{2M}{r}}},
\end{equation}
and they are suppressed as $r^{-3/2}$ when $r$ increases. Since we intend to apply the quantum tunneling technique to investigate the possible existence of Hawking radiation, we are interested in how Eq.~\eqref{eq:commutator} scales when approaching the event horizon in the semiclassical limit, i.e., when $r\to2M$ for large $M$. Taking the limit explicitly yields
\begin{equation} \label{eq:commutatorNH}
    k_r\left.\left[k_r,\kappa(r)\right]\right|_{r\to2M} \propto k_r\ M^{-1}.
\end{equation}
The remaining terms in the eigenvalue equation are quadratic in the momentum components. Hence, neglecting the commutator terms amounts to assuming $|k_r|M^{-1} \ll k_i^2$, where $i$ denotes the components. This is satisfied when $\kappa(r)$ varies slow enough in space. Thus, the commutator contributions to the eigenvalues are subleading, and since we are focusing on the leading-order contribution to analogue Hawking radiation, they are neglected in this treatment.

Here, however, we must stress that we are not assuming that $\frac{\partial \kappa(r)}{\partial _r}=0$ at the horizon. We are merely pointing out that the contributions from such terms in the eigenvalue equation are subleading. The dynamics of $\kappa(r)$ at $r=2M_0$ are essential for the existence of the analogue horizon, and its meaning will become apparent in Secs.~\ref{sec:experiment} and \ref{sec:conclusions}.

\subsection{Particle and antiparticle contributions near the horizon} \label{sec:channels}

In the quantum tunneling method, there are two processes that give massless contributions to the spontaneous black hole emission: a particle tunneling outwards from the interior of the black hole, and an antiparticle tunneling into the black hole. The quantity of interest describing classically forbidden trajectories is the imaginary part of the action of a radially infalling light-like shell of energy \cite{PW2000}. As anticipated, we consider the region in momentum space corresponding to the EC and since the spatial dependence of the tilting term is taken to be in the $x$-coordinate, we solve Eq.~\eqref{eq:exconekz}, which are the eigenvalues of the model in Eq.~\eqref{eq:Dirac op}, for $k_x$. Recalling the identifications $x:=r$ and $\kappa := +\sqrt{\frac{2M}{r}}$, we find two solutions
\begin{equation} \label{eq:momentumbranch}
    k_r^{\pm} = \frac{-\sqrt{\frac{2M}{r}} k_z}{\frac{2M}{r}-1} \pm \sqrt{\frac{k_z^2}{\left(\frac{2M}{r}-1\right)^2}+\frac{k_y^2}{\frac{2M}{r}-1}}.
\end{equation}
Here, the expression inside the square-root is bound to be positive, since $k_r^{\pm}$ has to be real, such that $k_z^2 \geq - (\frac{2M}{r}-1)k_y^2$.

To determine which of these two solutions is to be interpreted as the momentum of the analogue particles and antiparticles, we investigate the behavior of Eq.~\eqref{eq:momentumbranch} when approaching asymptotic infinity, i.e., when $r\to +\infty$, in close analogy with the discussion on geodesics presented in Sec.~\ref{sec:BHrad}. In this limit, we find
\begin{align} \label{eq:asinf}
    \lim_{r\to +\infty} k_r^{\pm} &= \pm \sqrt{k_z^2-k_y^2},
\end{align}
with $k_z^2 \geq k_y^2$.
Consequently, $k_r^+\geq0$ and $k_r^-\leq 0$ when $r\to +\infty$. Recalling how the radial null geodesics $\dot{r}^{\pm}$, and thus $p_r^{\pm}$, which is the momentum conjugate to $r$, behaves at asymptotic infinity \cite{PW2000}, we see that $k_r^{\pm}$ shares the behavior of $p_r^{\pm}$. Hence, we interpret $k_r^{\pm}$ as analogue to the conjugate momentum $p_r$ meaning that $k_r^+$ and $k_r^-$ describe the momentum of an analogue particle and antiparticle, respectively. 

Having identified the different momentum channels, we now turn to investigate the presence of classically forbidden processes. These are described by the imaginary part of the action, which for an analogue radially infalling massless shell of energy reads \cite{Volovik1999, PW2000},
\begin{equation} \label{eq:fullact}
    \text{Im}\left(I\right) =  \text{Im}\left(\int dr \, k_r \right),
\end{equation}
which is to be evaluated on the EC. Formally, the actions for each of the channels on the EC are given by
\begin{align} \label{eq:fullact2}
    \text{Im}\left(I^{\pm}\right) &=\text{Im}\left\{ \int dr \times \right. \nonumber
    \\
    &\quad\left.\left[ \frac{-\sqrt{\frac{2M}{r}} k_z}{\frac{2M}{r}-1} \pm \sqrt{\frac{k_z^2}{\left(\frac{2M}{r}-1\right)^2}+\frac{k_y^2}{\frac{2M}{r}-1}}\right]\right\},
\end{align}
and lead to
\begin{align}
    \text{Im}\left[I^+(k_z)\right] &= \begin{cases} 2\pi M|k_z| \quad &k_z <0 \\ 0 \quad &k_z \geq 0, \end{cases} \label{eq:imactpfinal2}
    \\
    \text{Im}\left[I^-(k_z)\right] &= \begin{cases} 2\pi M|k_z| \quad &k_z >0 \\ 0 \quad &k_z \leq 0, \end{cases}\label{eq:imactapfinal2}
\end{align}
where detailed calculations are presented in Appendix~\ref{app:imact}.
The nonzero contributions to the imaginary part of the action indicate that classically forbidden processes reminiscent of those happening near a black hole horizon are present in our model. It should be noted that Eqs.~\eqref{eq:imactpfinal2} and \eqref{eq:imactapfinal2} are independent of $k_y$, which consequently does not affect the classically forbidden processes.

\subsection{Semiclassical emission rate} \label{sec:emission}

The probability of a classically forbidden process is estimated by the semiclassical quantum tunneling rate. Since Hawking radiation corresponds to both particles tunneling out of and antiparticles tunneling into the black hole, both channels contribute to the total probability of the tunneling process. The associated tunneling rate can then be interpreted as the analogue black hole emission rate as long as the respective tunneling amplitudes are added up and then squared \cite{PW2000}. Using the imaginary parts of the actions derived in Sec.~\ref{sec:channels}, the respective amplitudes read
\begin{align}
    \mathcal{A}^+ &=e^{-2\pi M |k_z|}\theta(-k_z), \label{eq:amplitude_particle}
    \\
    \mathcal{A}^- &=e^{-2\pi M |k_z|}\theta(k_z), \label{eq:amplitude_antiparticle}
\end{align}
where $\theta(k_z)$ is the Heaviside theta-function defined as
\begin{equation}
    \theta(x) = \begin{cases} 1 \quad &x>0 \\0 \quad &x\leq 0. \end{cases}
\end{equation}
From the amplitudes, the semiclassical emission rate reads
\begin{equation} \label{eq:emission1}
    \Gamma\sim \left(\mathcal{A}^++\mathcal{A}^-\right)^2=\left(\mathcal{A}^+\right)^2+\left(\mathcal{A}^-\right)^2 + \mathcal{A}^+\mathcal{A}^-.
\end{equation}
Since $\text{Im}\left(I^+\right)$ and $\text{Im}\left(I^-\right)$ do not have support on each others domains [cf. Eqs.~\eqref{eq:amplitude_particle} and \eqref{eq:amplitude_antiparticle}], the cross term $\mathcal{A}^+\mathcal{A}^-$ vanishes. We thus find
\begin{equation} \label{eq:emission}
    \Gamma \sim e^{-4\pi M |k_z|} \qquad k_z \neq 0,
\end{equation}
which estimates the total probability for a tunneling process to occur. Given the interpretation of $\mathcal{H}$ in Eq.~\eqref{eq:NHham} as an analogue model for a black hole, the quantity in Eq.~\eqref{eq:emission} can be thought of as an emission rate, the nature of which is system dependent. Note that while the probability is finite and non-zero for $|k_z|\neq 0$, it is exponentially suppressed with increasing values of $|k_z|$. This means that a tunneling process is unlikely unless $|k_z|$ is small.

\section{Towards Experimental Realizations} \label{sec:experiment}

Here, we connect our model to possible experimental realizations in order to specify the concrete nature of the analogue Hawking radiation. One promising experimental platform for realizing our model is in photonic crystals, where gain and loss can be readily implemented and which are well described by NH Hamiltonians \cite{Ozdemir2019, bfermiarcs, NHreview, NHPC}.

The Hermitian part of our NH Hamiltonian $\mathcal{H}$ in Eq.~\eqref{eq:NHham} looks like a Weyl Hamiltonian, which has been previously realized in photonic crystals in both the Hermitian and non-Hermitian form \cite{bfermiarcs, Chen2016, LWYRFJS2015, Yang2017}. The anti-Hermitian part of $\mathcal{H}$ provides a bigger challenge for realization. While its momentum-dependent imaginary terms can be realized using single-photon interferometry \cite{expknots}, we are not aware of a realization in a photonic crystal. Let us suggest a possible path towards realizing such terms. We start by coupling the photonic crystal to its environment in such a way that the dynamics of the photonic lattice are well-described by the Lindblad master equation \cite{Lindblad1976},
\begin{equation}
    \partial_t \rho = i (\rho H_\textrm{eff}^\dagger - H_\textrm{eff} \rho)  + \sum_n( \kappa L_{x,n} \rho L_{x,n}^\dagger + L_{z,n} \rho L_{z,n}^\dagger),
\end{equation}
where $\rho$ is the density matrix, $L_{i,n}$ are the jump operators, $H_\textrm{eff}$ is the effective non-Hermitian Hamiltonian which yields
\begin{equation}
  H_\textrm{eff} = H_\textrm{H} - i \sum_{n}( \kappa L^\dagger_{x,n} L_{x,n} +  L^\dagger_{z,n} L_{z,n})/2,
\end{equation}
with $H_\textrm{H}= \sum_{\bf k} {\bf b}^\dagger_{\bf k} \mathcal{H}_\textrm{H} {\bf b}_{\bf k}$ the (Hermitian) system Hamiltonian, where ${\bf b}_{\bf k} = (b_{1,\bf k}, b_{2,\bf k})^T$ with $b_{i,\bf k}$ annihilating a particle at site $i$ with momentum ${\bf k}$, and $\sum_n( \kappa L_{x,n} \rho L_{x,n}^\dagger + L_{z,n} \rho L_{z,n}^\dagger)$ is known as the quantum jump or recycling term. This assumption, namely that the bath acts as a Markovian bath, is well justified for the majority of optical experiments \cite{Daley2014}.
As the photonic lattice is effectively a classical system, the quantum jump terms do not contribute and the dynamics from the master equation coincide with the dynamics generated by the effective non-Hermitian Hamiltonian $H_\textrm{eff}$ \cite{Ozdemir2019, bfermiarcs, NHreview, NHPC,Ozawa2019}. We are thus left with having to engineer the correct jump operators such that $\mathcal{H}_\textrm{eff}$ in $H_\textrm{eff}= \sum_{\bf k} {\bf b}^\dagger_{\bf k} \mathcal{H}_\textrm{eff} {\bf b}_{\bf k}$ resembles $\mathcal{H}$ in Eq.~\eqref{eq:NHham}.

If the photonic crystal whose system Hamiltonian is modelled by $\mathcal{H}_\textrm{H} = k_x \sigma^x + k_y \sigma^y$, i.e., the Hermitian part of $\mathcal{H}$ \eqref{eq:NHham}, is coupled to the environment such that its jump operators read
\begin{equation}
   L_{x,n} = b_{2,n} - i b_{2, n + \hat{x}}\ \  \text{and}\ \ L_{z,n} = b_{2, n} - i b_{2, n + \hat{z}}, \label{eq:non_local_jump_op}
\end{equation}
where $b_{2, n+\hat{e}}$ is a real-space annihilation operator at site $2$ in unit cell $n+\hat{e}$, we find that the effective Fourier-transformed Hamiltonian in the linearized form reads
\begin{equation}
\mathcal{H}_\textrm{eff} = k_x\sigma^x+k_y\sigma^y  - i (2 + \kappa k_x+ k_z ) (\sigma_0 - \sigma_z),
\end{equation}
(see Appendix~\ref{app:der_eff_ham} for a derivation of this effective Hamiltonian). This effective description has eigenvalues \begin{equation}
E_{\textrm{eff},\pm} = - i (2 + k_z + \kappa k_x) \pm \sqrt{k_x^2 + k_y^2 - (2 + k_z + \kappa k_x)^2},
\end{equation}
from which we see that the effective Hamiltonian thus indeed realizes ECs albeit at a shifted $k_z$ value $(k_z \rightarrow k_z - 2)$.
Nonlocal jump operators $L_{i,n}$ as in Eq.~\eqref{eq:non_local_jump_op} have been realized in optical cavity arrays through the adiabatic elimination of a third auxiliary cavity coupled to two optical cavities \cite{Metelmann2015, Bernier2017, Malz2018}.

As we pointed out in Sec.~\ref{sec:commutator}, the non-trivial spatial dynamics of $\kappa(r)$ are essential for the existence of a horizon. This ensures that the relevant system, in this particular case the photonic crystal, hosts two separate regions reminiscent of the interior and the exterior of a black hole. The horizon is not the only consequence of this feature, however. In fact, the non-trivial dynamics of $\kappa(r)$ also act as a source of the analogue Hawking radiation. We already showed that $\frac{\partial\kappa(r)}{\partial r}$ is proportional to the mass of the analogue black hole, cf. Eq.~\eqref{eq:commutator}, which in the case of a Schwarzschild black hole is related to the Hawking temperature $T_{\text{BH}}$, cf. Eq.~\eqref{eq:Hawkingtemp}. Explicitly, the temperature is given by
\begin{equation}
    2\pi T_{\text{BH}} = -\left.\frac{\partial \kappa(r)}{\partial r} \right|_{r=2M_0}.
\end{equation}
This means that the rate at which the analogue black hole radiates can be controlled by tuning the behavior of $\kappa(r)$ around the analogue horizon. Note, however, that the change of $\kappa(r)$, and hence the temperature $T_{\text{BH}}$, must be small in order for the previous calculations to be valid.

Let us now discuss how the features of Hawking radiation manifest in an actual setup and what constraints one should pose on such a photonic crystal. As the important feature is the existence of an analogue horizon separating two regions reminiscent of the interior and the exterior of the black hole, respectively, $\kappa$ has to vary between $\kappa_1<1$ and $\kappa_2>1$. This variation has to be in the $x:=r$ direction according to Sec.~\ref{sec:cones}, such that $\kappa_1<\kappa<\kappa_2$. Thus, the horizon will take the form of a plane in the $x$-direction of the photonic crystal. The analogue Hawking radiation in such a system stems from quantum tunneling of entangled photons through the analogue horizon (where $\kappa=1$). After the tunneling process, these particles are free to move through the analogue spacetime --- anti-particles in the region where $\kappa>1$, and particles in the region where $\kappa<1$. Eventually, some of these entangled particles may exit the photonic crystal, and thus serve as a signature of analogue Hawking radiation. We, however, recall that photonic crystals are known to emit thermal radiation \cite{Lin2003,Florescu2005}, meaning that the entangled photons have to be distinguished from those comprising the (uncorrelated) thermal radiation. Despite providing an experimental challenge, both in constructing systems with the correct tilting dynamics and in making appropriate measurements, this provides all the ingredients needed to look for signatures of analogue Hawking radiation.

\section{Conclusions and Outlook}\label{sec:conclusions}

In this work, we study analogies between the band structure of a $\mathcal{PT}$-symmetric NH two-band model and curved spacetime geometries. As an extension of analogies between Weyl cones and spacetime light cones, we find a relation between the $\mathcal{PT}$ symmetry-protected cone of EPs, dubbed the EC, and the light cone of a radially infalling observer in the presence of a (3+1)D Schwarzschild black hole. By calculating the semiclassical emission rate, we show that radiative processes reminiscent of those happening in black holes are present in the NH model. We show that this particular model can be retrieved as an effective description originating from a Lindblad master equation describing the dynamics of a photonic crystal. In this context, we propose a concrete setup within experimental reach in which the analogue Hawking radiation might be observed.

We note that our toy model in Eq.~\eqref{eq:NHham} not only hosts ECs, but that the eigenvalues themselves also form cones. Indeed, both the real and imaginary parts of the eigenvalues in Eq.~\eqref{eq:NHham} display cone structures, which are separated from each other in the complex energy plane by EPs. By tuning $\kappa$, these cones do not tilt unlike the ECs, but instead rotate in the energy plane. However, by adding an additional ($\mathcal{PT}$ symmetry-preserving) term to our Hamiltonian, e.g., $\kappa_1 k_x \sigma^0$, these cones can tilt similar to the spacetime light cones upon tuning $\kappa_1$. In fact, this is a generic feature of $\mathcal{PT}$-symmetric NH models in three dimensions that consists of at most linear terms in momentum. While we thus set out in this paper to investigate the connection between ECs and spacetime light cones (due to the attractive similarities between the two beyond those between Hermitian Weyl cones and light cones), a natural additional question to ask is whether such a connection can also exist between these NH eigenvalue cones and light cones. We note that to investigate this question, the method we have employed in this paper may not be applicable. At this point, it is not clear how the vielbein construction can be extended to NH Dirac-like operators, which explicitly include imaginary terms, and we leave this question for further studies.

Even though the analogy constructed in this work is an extension of the previously known Hermitian analogies, it is fundamentally different. This is especially clear in the interpretation of the analogue Hawking radiation. In particular, the tilting parameter $\kappa(r)$ comprise an imaginary term in the Hamiltonian $\mathcal{H}$, and is therefore intimately related to dissipative features. Furthermore, our calculations indicate a relation between the dynamical behavior of these anti-Hermitian terms, and a pair-creation mechanism. This connection has, to our knowledge, not been made before. Combining these conclusions, the analogy suggests a novel way to model realistic astrophysical black holes, which are massive (large $M$) and hence, cold (small $T_{\text{BH}}$). This is furthermore consistent with the semiclassical treatment, which motivated neglecting subleading powers of $M$ in Eq.~\eqref{eq:commutator}. A natural continuation along these lines is, however, to investigate the meaning of these subleading corrections. We note that these amount to commutator contributions to the eigenvalue equation and thus the equation of the EC, Eq.~\eqref{eq:exconekz}. Keeping those terms would make the relation to the analogue metric Eq.~\eqref{eq:analmetricexam} less apparent, as Eq.~\eqref{eq:exconekz} could host imaginary terms. Therefore, studies and interpretations of such corrections are left for future works.


Another interesting quantity that could be investigated further, apart from the previously computed emission rate, is the entropy. One could calculate the entanglement entropy associated to the tilted NH Weyl-like Hamiltonian and see if it matches the expected result of the Bekenstein-Hawking entropy for the analogue semiclassical black hole \cite{Bekenstein,Entanglemententropy}. As the present analogy focuses on finding a meaningful analogue of Hawking radiation, such a computation goes beyond the scope of this particular study and is left for future works.

The present study investigates analogue Hawking radiation on the EC, where $\mathcal{PT}$ symmetry is spontaneously broken. One could wonder if further generalizations of this analogy can lead to more abstract connections between ECs and analogue event horizons, where the $\mathcal{PT}$ phase transition may be related to the transition between the time-like (exterior) and space-like (interior) regions of the black hole experienced by an infalling observer crossing the black hole horizon. Furthermore, it is interesting to point out that the interior of the EC satisfying $E^2<0$ can be identified as a tachyon-like phase, where excitations travel faster than the effective speed of light \cite{Feinberg1967}. We note that such a connection between $\mathcal{PT}$-symmetric models and tachyonic physics was the focus of a very recent study, where a tachyonic phase was identified in the phase diagram of a 1D Dirac model with real and imaginary mass terms \cite{Sticlet2021}. However, to our knowledge, connections between spontaneous $\mathcal{PT}$ symmetry-breaking and horizon physics have not been investigated enough to establish any precise relations. We nevertheless note that in the context of gauge-gravity duality, recent work has made concrete connections between $\mathcal{PT}$-symmetric quantum theory and black holes in asymptotic AdS spacetimes \cite{NHholography}. A possible intriguing relation between these ideas realized on holographic setups dual to $\mathcal{PT}$ phase transitions and our work is left for further studies.

Lastly, the importance of multiband models in similar NH models has been studied in recent works \cite{3EPemil,HOEP,4EPMarcus}. In particular, the appearance of higher-order EPs is investigated. To this point, we note that considering additional bands, the ECs of order two can be accompanied by third-order exceptional lines and fourth-order EPs in 3D $\mathcal{PT}$-symmetric systems \cite{4EPMarcus}. The impact of the higher-order exceptional structure  on the physics discussed in this paper may be interesting but is outside the scope of this work. Here we assume additional bands to be located far away from the ones considered in this work, and leave further investigation for future studies.\\

\emph{Note added:} After the publication of this paper, we have become aware of Ref.~\onlinecite{Sundaram2021}, where a connection is made between $\mathcal{PT}$-symmetry breaking transitions and the inverse-square potential relevant for black hole physics near the horizon. The content of that work thus forms another motivation to search for connections between near-horizon black hole physics and $\mathcal{PT}$-symmetric systems. \\

\section*{Acknowledgments}
We would like to thank Emil J. Bergholtz, Johan Carlstr\"{o}m and Watse Sybesma for useful discussions. Special thanks to Thors Hans Hansson, Fawad Hassan, Bo Sundborg and Lárus Thorlacius for useful comments on the manuscript. M.S. is supported by the Swedish Research Council (VR) and the Wallenberg Academy Fellows program of the Knut and Alice Wallenberg Foundation. L.R. is supported by the Knut and Alice Wallenberg foundation under grant no. 2017.0157. F.K.K. is supported by the Max Planck Institute of Quantum Optics (MPQ) and Max-Planck-Harvard Research Center for Quantum Optics (MPHQ). The work was finished while J.L.A. and L.R. were participating in a mathematical physics conference, organized by Uppsala University and funded by the Göran Gustafsson foundation, whose stimulating environment is also acknowledged.

 \newpage
\begingroup
\onecolumngrid
\appendix
\section{Cones of exceptional points in $\mathcal{P}\mathcal{T}$-symmetric systems and their analogue event horizons} \label{app:conerelation}
In this Appendix, we show that any 3D NH $\mathcal{PT}$-symmetric two-band model linear in momentum host cones of EPs, which can be mapped to a curved spacetime metric. Since we are only interested in the EPs, and not the eigenvalues themselves, we can set the term proportional to the identity matrix to $0$. Generally, in an appropriate choice of coordinates, the Hamiltonian attains the form
\begin{align} \label{eq:appHam}
    H &= \left(a_xk_x+a_yk_y+a_0\right)\sigma^x+\left(b_xk_x+b_yk_y+b_0\right)\sigma^y+i\left(c_xk_x+c_yk_y+c_zk_z+c_0\right)\sigma^z,
\end{align}
with all parameters being real. We note that the constant terms $a_0$, $b_0$ and $c_0$ can be set to $0$, since they can be removed by shifting the momentum components. Explicitly, the corresponding EPs are given by
\begin{align} \label{eq:excone}
    &\quad \left(a_x^2+b_x^2-c_x^2\right)k_x^2 + \left(a_y^2+b_y^2-c_y^2\right)k_y^2-c_z^2k_z^2 +2k_xk_y\left(a_xa_y+b_xb_y-c_xc_y\right)-2c_yc_zk_yk_z-2c_xc_zk_xk_z=0,
\end{align}
and solving for $k_z$ yields
\begin{align}
    k_z &=-\frac{1}{c_z}\left(c_xk_x+c_yk_y\right)\pm \sqrt{\left(a_xk_x+a_yk_y\right)^2+\left(b_xk_x+b_yk_y\right)^2}.
\end{align}
with $c_z\neq 0$. As in the simple example dealt with in the main text, $k_z$ can be thought of as the eigenvalue of some Dirac-like operator, in this case, e.g.,
\begin{equation} \label{eq:appDop}
     \hat{k}_z = -\frac{1}{c_z}\left(c_xk_x+c_yk_y\right)\sigma^0+\left(a_xk_x+a_yk_y\right)\sigma^x+\left(b_xk_x+b_yk_y\right)\sigma^y,
\end{equation}
whose eigenvalues form a cone. It should be stressed that adding a term proportional to $\sigma^z$ in Eq.~\eqref{eq:appDop} would correspond to adding an equivalent term to the initial Hamiltonian in Eq.~\eqref{eq:appHam}, which would break the implemented $\mathcal{PT}$-symmetry. Hence, such terms are forbidden, and we note more accurate that $k_z$ can be thought of as the eigenvalues of a Hermitian Dirac-operator subject to $\mathcal{PT}$ symmetry. 

Since a Dirac-operator can be expressed in terms of vielbeins \cite{Volovik2014,Horova2005}, we can write $\hat{k}_z$ on the form
\begin{equation}
    \hat{k}_z=e\indices{^i_a}k_i\sigma^a+e\indices{^i_0}k_i\sigma^0,
\end{equation}
with
\begin{align}
    e\indices{^0_0}&=1, \quad e\indices{^x_0}=-\frac{c_x}{c_z}, \quad e\indices{^y_0}=-\frac{c_y}{c_z}, \quad e\indices{^0_x}=0, \nonumber
    \\
    e\indices{^0_y}&=0, \quad e\indices{^x_x}=a_x,\quad e\indices{^x_y} = b_x,\quad e\indices{^y_x}=a_y, \quad e\indices{^y_y}=b_y,
\end{align}
from which a curved spacetime metric can be defined through $g_{\mu\nu}= e\indices{_{\mu}^{\alpha}}e\indices{_{\nu}^{\beta}}\eta_{\alpha\beta}$.

\section{Imaginary parts of the actions: Detailed calculations} \label{app:imact}
\begin{figure*}[hbt!]
\centering
\includegraphics[scale=0.2]{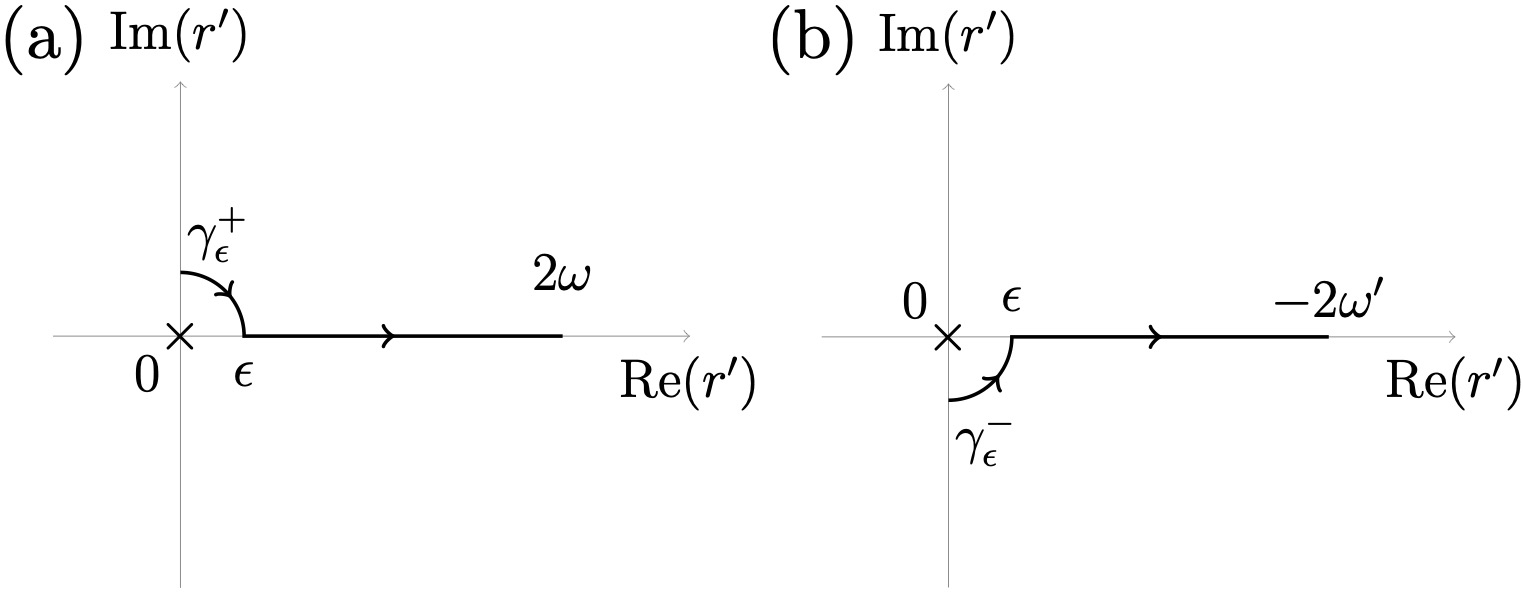}
\caption{The contours used when integrating (a) Eq.~\eqref{eq:appI+} and (b) Eq.~\eqref{eq:appI-}, respectively. The contours along the real line has to be deformed differently to reflect the correct physics of particles and antiparticles. Singularities in the description of particles are pushed to the lower half complex plane, and hence the contour is deformed to a quarter circle in the first quadrant in (a). Similarly, the singularities in the description of antiparticles are pushed to the upper half complex plane, giving a quarter circle in the fourth quadrant (b).}\label{fig:Contour}
\end{figure*}
In this Appendix, we provide the details for the calculation of the imaginary parts of the actions of the particle and antiparticle channels. The imaginary part of the action of the analogue radially infalling shell of energy \cite{PW2000} reads 
\begin{equation} \label{eq:appfullact}
   \text{Im}\left( I\right) =  \text{Im}\left(\int k_r dr\right),
\end{equation}
which we want to evaluate on the EC. Formally, the actions for each of the channels on the EC are given by
\begin{align}
   \text{Im}\left( I^{\pm}\right) &= \text{Im}\left\{\int dr \left[ \frac{-\sqrt{\frac{2M}{r}} k_z}{\frac{2M}{r}-1} \pm \sqrt{\frac{k_z^2}{\left(\frac{2M}{r}-1\right)^2}+\frac{k_y^2}{\frac{2M}{r}-1}}\right]\right\}.
\end{align}
We know that these solutions correspond to the momentum of a particle $(+)$ and antiparticle $(-)$, respectively. For the momentum of a particle, we consider a spontaneous radiation process in which a positive energy $\omega>0$ is assumed to radiate out from the interior of the black hole such that the mass of the black hole changes from $M$ to $M-\omega$. Thus, the action of the particle channel reads
\begin{align} \label{eq:apppartaction}
\text{Im}\left(I^+\right)&=\text{Im}\left\{\int_{2M}^{2\left(M-\omega\right)}dr  \left[\frac{-\sqrt{\frac{2M}{r}} k_z}{\frac{2M}{r}-1} + \sqrt{\frac{k_z^2}{\left(\frac{2M}{r}-1\right)^2}+\frac{k_y^2}{\frac{2M}{r}-1}}\right]\right\}.
\end{align}
The antiparticle case, on the other hand, can be interpreted as a particle with a negative energy $\omega'<0$ tunneling inwards such that the mass changes from $M$ to $M+\omega'$. Therefore, the action of the antiparticle channel is.
\begin{align} \label{eq:appapartaction}
 \text{Im}\left(I^-\right)&=\text{Im}\left\{\int_{2M}^{2\left(M+\omega'\right)} dr  \left[\frac{-\sqrt{\frac{2M}{r}} k_z}{\frac{2M}{r}-1} - \sqrt{\frac{k_z^2}{\left(\frac{2M}{r}-1\right)^2}+\frac{k_y^2}{\frac{2M}{r}-1}}\right]\right\}.
\end{align}
Re-writing and making a change of variables, $-r'=r-2M$, $dr=-dr'$, gives
\begin{align}
    \text{Im}\left(I^+\right) &= \text{Im}\left\{\int_0^{2\omega} dr' \left[\frac{\sqrt{2M\left(2M-r'\right)}k_z}{r'} -\sqrt{\frac{k_z^2\left(2M-r'\right)^2}{r'^2}+\frac{k_y^2\left(2M-r'\right)}{r'}}\right]\right\}, \label{eq:appI+}
    \\
   \text{Im}\left( I^-\right) &= \text{Im}\left\{\int_0^{-2\omega'} dr'\left[\frac{\sqrt{2M\left(2M-r'\right)}k_z}{r'} +\sqrt{\frac{k_z^2\left(2M-r'\right)^2}{r'^2}+\frac{k_y^2\left(2M-r'\right)}{r'}}\right]\right\}. \label{eq:appI-}
\end{align}
When evaluating these integrals, we have to proceed with caution. First, note that there are branch points at $r'=2M:=r'_1$, which is positive, and $r'= 2M \frac{k_z^2}{k_z^2-k_y^2}:=r'_2$, which is negative when $k_z^2< k_y^2$, and positive when $k_z^2> k_y^2$. Thus, when $k_z^2< k_y^2$, we make branch cuts from $r'=r_1'$ to positive infinity, and from $r'=r'_2$ to negative infinity. When $k_z^2>k_y^2$, we first note that $r_2'\geq r_1'$, and hence we make a branch cut from $r'=r_1'$ to $r'=r_2'$. When $k_z^2=k_y^2$, $r_2'\to \pm \infty$, depending on if $k_z^2\to k_y^2$ from above $(+)$ or below $(-)$. These cases are thus included in the cases above. 

Noting that in both $\omega\leq M$ and $-\omega'\leq M$, the energy of the radiated quanta cannot be larger than the initial mass of the black hole, the square root functions are analytic on the interval that is integrated over. Second, both integrals start at a pole of order 1, and evaluation of the corresponding integrals requires a deformation of the integral contour. Since $I^+$ and $I^-$ are actions of particles and antiparticles, respectively, the corresponding deformations are necessarily different as required by causality; the poles of the particle action $I^+$ are to be located in the lower half complex $r'$-plane, while the poles of the antiparticle action $I^-$ are to be located in the upper half complex $r'$-plane. Therefore, the contour of integration in $I^+$ is deformed to a quarter circle of radius $\epsilon$ in the first quadrant, connected to a line from $\epsilon$ to $2\omega$, while the contour of integration in $I^-$ is deformed to a quarter circle of radius $\epsilon$ in the fourth quadrant, connected to a line from $\epsilon$ to $-2\omega'$. The appearance of the different contours are displayed in Fig.~\ref{fig:Contour}. By taking the limit $\epsilon\to 0$, the original integrals are obtained. Hence,
\begin{align}
    \text{Im}\left(I^+\right) &= \text{Im}\left\{\lim_{\epsilon\to 0}\left(\int_{\gamma^+_{\epsilon}} + \int_{\epsilon}^{2\omega}\right)dr' \left[\frac{\sqrt{2M\left(2M-r'\right)}k_z}{r'} -\sqrt{\frac{k_z^2\left(2M-r'\right)^2}{r'^2}+\frac{k_y^2\left(2M-r'\right)}{r'}}\right]\right\},
    \\
    \text{Im}\left(I^-\right) &= \text{Im}\left\{\lim_{\epsilon\to 0}\left(\int_{\gamma^-_{\epsilon}} + \int_{\epsilon}^{-2\omega'}\right)dr' \left[\frac{\sqrt{2M\left(2M-r'\right)}k_z}{r'} +\sqrt{\frac{k_z^2\left(2M-r'\right)^2}{r'^2}+\frac{k_y^2\left(2M-r'\right)}{r'}}\right]\right\},
\end{align}
where $\gamma_{\epsilon}^+$ and $\gamma^-_{\epsilon}$ denote the quarter circles of radii $\epsilon$ located in the first and fourth quadrant, respectively [cf. Fig~\ref{fig:Contour}]. The second integrals integrate over a domain where the corresponding integrand is real valued. Hence, they do not contribute to the imaginary parts and therefore we only evaluate the integrals over $\gamma_{\epsilon}^+$ and $\gamma^-_{\epsilon}$, respectively. Choosing a parametrization as $r'=\epsilon e^{i\phi}$, $dr'=i\epsilon e^{i\phi} d\phi$, where $\phi\in (-\pi,\pi]$ is the argument of $r'$ in the principal branch, the imaginary parts of the actions read
\begin{align}
    \text{Im}\left(I^+\right) &= \lim_{\epsilon \to 0} \int_{\pi/2}^0 d\phi \left[ \sqrt{2M \left( 2M-\epsilon e^{i\phi}\right)}k_z  -\sqrt{k_z^2\left( 2M - \epsilon e^{i\phi}\right)^2+\epsilon e^{i\phi}\left(2 M-\epsilon e^{i\phi}\right)k_y^2}\right],
    \\
    \text{Im}\left(I^-\right) &= \lim_{\epsilon \to 0} \int_{-\pi/2}^0 d\phi \left[ \sqrt{2M \left( 2M-\epsilon e^{i\phi}\right)}k_z +\sqrt{k_z^2\left( 2M - \epsilon e^{i\phi}\right)^2+\epsilon e^{i\phi}\left(2 M-\epsilon e^{i\phi}\right)k_y^2}\right].
\end{align}
Since the integrands are bounded from above in the limit $\epsilon \to 0$, the order of the limit and the integration can be interchanged. Doing so, we finally arrive at
\begin{align}
    \text{Im}\left[I^+(k_z)\right] &= \begin{cases} 2\pi M|k_z| \quad &k_z <0 \\ 0 \quad &k_z \geq 0, \end{cases} \label{eq:appimactpfinal2}
    \\
    \text{Im}\left[I^-(k_z)\right] &= \begin{cases} 2\pi M|k_z| \quad &k_z >0 \\ 0 \quad &k_z \leq 0. \end{cases}\label{eq:appimactapfinal2}
\end{align}

\section{Derivation of the effective Hamiltonian} \label{app:der_eff_ham}

Here we derive the effective Hamiltonian $H_\textrm{eff}$ that appears in the Lindblad master equation in Sec.~\ref{sec:experiment}.
We couple the photonic crystal to its environment with non-local jump operators $L_{x,n} = b_{2,n} - i b_{2, n + \hat{x}}$ and $L_{z,n} = b_{2, n} - i b_{2, n + \hat{z}}$, where $b_{2, n+\hat{e}}^\dagger$ ($b_{2, n+\hat{e}}$) is a real-space creation (annihilation) operator at site $2$ in unit cell $n+\hat{e}$. We then find that 
\begin{equation}
\sum_{n}( \kappa L^\dagger_{x,n} L_{x,n} +  L^\dagger_{z,n} L_{z,n}) = 2  \sum_{\bf q}  (2 + \kappa \sin{q_x}+ \sin{q_z} ) b^\dagger_{2, \bf q} b_{2, \bf q},
\end{equation}
where we have performed a Fourier transform $b_n = \sum_{\bf q} b_{\bf q} e^{i {\bf q} \cdot {\bf r}_n}$. Expanding ${\bf q}$ around zero, this term reads $2  \sum_{\bf k}  (2 + \kappa k_x+ k_z ) b^\dagger_{2, \bf k} b_{2, \bf k}$, where ${\bf q} \approx {\bf k}$. We then find $H_\textrm{eff}= \sum_{\bf k} {\bf b}^\dagger_{\bf k} \mathcal{H}_\textrm{eff} {\bf b}_{\bf k}$ with ${\bf b}_{\bf k} = (b_{1,\bf k}, b_{2,\bf k})^T$ and
\begin{equation}
\mathcal{H}_\textrm{eff} = k_x\sigma^x+k_y\sigma^y  - i (2 + \kappa k_x+ k_z ) (\sigma_0 - \sigma_z).
\end{equation}
This effective description has eigenvalues $E_{\textrm{eff},\pm} = - i (2 + k_z + \kappa k_x) \pm \sqrt{k_x^2 + k_y^2 - (2 + k_z + \kappa k_x)^2}$ from which we see that the effective Hamiltonian thus indeed realizes ECs albeit at a shifted $k_z$ value $(k_z \rightarrow k_z - 2)$.


\begin{thebibliography}{10}

\bibitem{QHE}
K.v. Klitzing, G. Dorda, and M. Pepper, {\em New Method for High-Accuracy Determination of the Fine-Structure Constant Based on Quantized Hall Resistance}, \href{http://dx.doi.org/10.1103/PhysRevLett.45.494}{Phys. Rev. Lett. {\bf 45}, 494 (1980)}.

\bibitem{hasankane}
M.Z. Hasan and C.L. Kane, {\em Colloquium: Topological insulators}, \href{http://dx.doi.org/10.1103/RevModPhys.82.3045}{Rev. Mod. Phys. {\bf 82}, 3045 (2010)}.

\bibitem{qizhang}
X.-L. Qi and S.-C. Zhang, {\em Topological insulators and superconductors}, \href{http://dx.doi.org/10.1103/RevModPhys.83.1057}{Rev. Mod. Phys. {\bf 83}, 1057 (2011)}.

\bibitem{goerbig}
M.O. Goerbig, {\em Electronic properties of graphene in strong magnetic field}, \href{http://dx.doi.org/10.1103/RevModPhys.83.1193}{Rev. Mod. Phys. {\bf 83}, 1193 (2011)}.

\bibitem{weylreview}
N.P. Armitage, E.J. Mele, and A. Vishwanath, {\em Weyl and Dirac semimetals in three-dimensional solids}, \href{http://dx.doi.org/10.1103/RevModPhys.90.015001}{Rev. Mod. Phys. {\bf 90}, 015001 (2018)}.

\bibitem{jantopreview}
J.C. Budich and B. Trauzettel, {\em From the adiabatic theorem of quantum mechanics to topological states of matter}, \href{http://dx.doi.org/10.1002/pssr.201206416}{physica status solidi (RRL) {\bf 7}, 109 (2013)}.

\bibitem{W1929} 
H. Weyl, {\em Elektron und Gravitation. I}, \href{http://dx.doi.org/10.1007/BF01339504}{Z. Physik {\bf 56}, 330 (1929)}.

\bibitem{XBAN2015} 
S.-Y. Xu, I. Belopolski, N. Alidoust, M. Neupane, G. Bian, C. Zhang, R. Sankar, G. Chang, Z. Yuan, C.-C. Lee, S.-M. Huang, H. Zheng, J. Ma, D.S. Sanchez, B. Wang, A. Bansil, F. Chou, P.P. Shibayev, H. Lin, S. Jia, and M.Z. Hasan, {\em Discovery of a Weyl fermion semimetal and topological Fermi arcs}, \href{http://dx.doi.org/10.1126/science.aaa9297}{Science {\bf 349}, 613 (2015)}.

\bibitem{LWFM2015}
B.Q. Lv, H.M. Weng, B.B. Fu, X.P. Wang, H. Miao, J. Ma, P. Richard, X.C. Huang, L.X. Zhao, G.F. Chen, Z. Fang, X. Dai, T. Qian, and H. Ding, {\em Experimental Discovery of Weyl Semimetal TaAs}, \href{http://dx.doi.org/10.1103/PhysRevX.5.031013}{Phys. Rev. X {\bf 5}, 031013 (2015)}.

\bibitem{LWYRFJS2015}
L. Lu, Z Wang, D. Ye, L. Ran, L. Fu, J. D. Joannopoulos, and M. Soljacic, {\em Experimental observation of Weyl points}, \href{http://dx.doi.org/10.1126/science.aaa9273}{Science {\bf 349}, 6248, 622-624 (2015)}.

\bibitem{Chiral1}
H. B. Nielsen and M. Ninomiya, {\em The Adler-Bell-Jackiw
anomaly and Weyl fermions in a crystal}, \href{http://dx.doi.org/10.1016/0370-2693(83)91529-0}{Phys. Lett. B
{\bf 130}, 6 (1983)}.

\bibitem{Chiral2}
K. Fukushima, D. E. Kharzeev, and H. J. Warringa,
{\em Chiral magnetic effect}, \href{http://dx.doi.org/10.1103/PhysRevD.78.074033}{Phys. Rev. D {\bf 78}, 074033
(2008)}.

\bibitem{Chiral3}
A. A. Zyuzin and A. A. Burkov, {\em Topological response
in weyl semimetals and the chiral anomaly}, \href{http://dx.doi.org/10.1103/PhysRevB.86.115133}{Phys. Rev.
B {\bf 86}, 115133 (2012)}.

\bibitem{Chiral4}
M. M. Vazifeh and M. Franz, {\em Electromagnetic Response
of Weyl Semimetals}, \href{http://dx.doi.org/10.1103/PhysRevLett.111.027201}{Phys. Rev. Lett. {\bf 111}, 027201 (2013)}.

\bibitem{Unruh}
W. G. Unruh, {\em Experimental black-hole evaporation?}, \href{https://journals.aps.org/prl/abstract/10.1103/PhysRevLett.46.1351}{Phys. Rev. Lett. {\bf 46}, 1351 (1981)}.

\bibitem{Barcelo}
C. Barcel\'o, S. Liberati, and M. Visser, {\em Analogue
gravity}, \href{https://link.springer.com/article/10.12942/lrr-2011-3}{Living Rev. Rel. {\bf 14}, 3 (2011)}.

\bibitem{beenakker}
C. Beenakker, Commentary at the Journal Club for Condensed Matter Physics (2015), \href{https://www.condmatjclub.org/?p=2644}{https://www.condmatjclub.org/?p=2644}.

\bibitem{Volovik2016}
G. E. Volovik, {\em Black hole and hawking radiation by type-II Weyl fermions}, \href{https://doi.org/10.1134/S0021364016210050}{JETP Letters  {\bf 104},  pages645–648 (2016)}.

\bibitem{zobkovblackholes}
M.A. Zubkov, {\em Analogies between the Black Hole Interior and the Type II Weyl Semimetals}, \href{http://dx.doi.org/10.3390/universe4120135}{Universe {\bf 4}, 135 (2018)}.

\bibitem{volovik}
G.E. Volovik, and K. Zhang, {\em Lifshitz Transitions, Type-II Dirac and Weyl Fermions, Event Horizon and All That}, \href{http://dx.doi.org/10.1007/s10909-017-1817-8}{J. Low Temp. Phys. {\bf 189}, (2017)}.

\bibitem{Zubkovold}
M.A. Zubkov, {\em The black hole interior and the type II Weyl fermions}, \href{https://www.worldscientific.com/doi/abs/10.1142/S0217732318500475}{Mod. Phys. Lett. A Vol. {\bf 33}, 1850047 (2018)}.

\bibitem{Liu2018}
H. Liu, J.-T. Sun, H. Huang, F. Liu, and S. Meng, {\em Fermionic analogue of black hole radiation
with a super high Hawking temperature}, \href{http://www.iop.cas.cn/xwzx/kydt/202012/P020201214554607990461.pdf}{Chin. Phys. Lett., {\bf 37}, 6, 067101 (2020)}.



\bibitem{Yaron2020} 
Y. Kedem, E.J. Bergholtz, and F. Wilczek {\em Black and White Holes at Matetrial Junctions}, \href{http://dx.doi.org/10.1103/PhysRevResearch.2.043285}{Phys. Rev. Research {\bf 2}, 043285 (2020)}.

\bibitem{Nissinen} 
J. Nissinen, G.E. Volovik, {\em Type-III and IV interacting Weyl points}, \href{http://dx.doi.org/10.1134/S0021364017070013}{Jetp Lett. {\bf 105}, 447–452 (2017)}.


\bibitem{Hawking1975}
S.W. Hawking, {\em Particle creation by black holes}, \href{http://dx.doi.org/10.1007/BF02345020}{Comm. Math. Phys. {\bf 43}, 199 (1975)}.




\bibitem{NHreview}
E.J. Bergholtz, J.C. Budich, and F.K. Kunst, {\em Exceptional topology of non-Hermitian systems}, \href{http://dx.doi.org/10.1103/RevModPhys.93.015005}{Rev. Mod. Phys {\bf 93}, 015995 (2021)}.




\bibitem{Lin2011}
Z. Lin, H. Ramezani, T. Eichelkraut, T. Kottos, H. Cao, and D. N. Christodoulides, {\em Unidirectional Invisibility Induced by PT-Symmetric Periodic Structures}, \href{https://journals.aps.org/prl/abstract/10.1103/PhysRevLett.106.213901}{Phys. Rev. Lett. {\bf 106}, 213901 (2011)}.

\bibitem{Regensburger2012}
A. Regensburger, C. Bersch, M.-A. Miri, G. Onishchukov, D. N. Christodoulides, and U. Peschel, {\em Parity–time synthetic photonic lattices}, \href{https://www.nature.com/articles/nature11298}{Nature {\bf 488}, 167 (2012)}.

\bibitem{Peng2014}
Bo Peng, \c{S}. K. \"{O}zdemir, F. Lei, F. Monifi, M. Gianfreda, G. L. Long, S. Fan, F. Nori, C. M. Bender, and L. Yang, {\em Parity–time-symmetric whispering-gallery microcavities}, \href{https://www.nature.com/articles/nphys2927}{Nat. Phys. {\bf 10}, 394 (2014)}.

\bibitem{Feng2014}
L. Feng, Z. J. Wong, R.-M. Ma, Y. Wang, and X. Zhang, {\em Single-mode laser by parity-time symmetry breaking}, \href{https://www.science.org/doi/10.1126/science.1258479}{Science {\bf 346}, 972 (2014)}.

\bibitem{Hodaei2014}
H. Hodaei, M.-A. Miri, M. Heinrich, D. N. Christodoulidesand, and M. Khajavikhan, {\em Parity-time–symmetric microring lasers}, \href{https://www.science.org/doi/10.1126/science.1258480}{Science {\bf 346}, 975 (2014)}.

\bibitem{Peng2016}
B. Peng, \c{S}. K. \"{O}zdemir, M. Liertzer, W. Chen, J. Kramer, H. Y{\i}lmaz, J. Wiersig, S. Rotter, and L. Yang, {\em Chiral modes and directional lasing at exceptional points}, \href{https://www.pnas.org/content/113/25/6845}{Proc. Natl. Acad. Sci. U.S.A. {\bf 113}, 6845 (2016)}.

\bibitem{Hodaei2017}
H. Hodaei, A. U. Hassan, S. Wittek, H. Garcia-Gracia, R. El-Ganainy, D. N. Christodoulides, and M. Khajavikhan, {\em Enhanced sensitivity at higher-order exceptional points}, \href{https://www.nature.com/articles/nature23280}{Nature {\bf 548}, 187 (2017)}.

\bibitem{Chen2017}
W. Chen, \c{S}. K. \"{O}zdemir, G. Zhao, J. Wiersig, and L. Yang, {\em Exceptional points enhance sensing in an optical microcavity}, \href{https://www.nature.com/articles/nature23281}{Nature {\bf 548}, 192 (2017)}.

\bibitem{Ozdemir2019}
\c{S}. K. \"{O}zdemir, S. Rotter, F. Nori, and L. Yang, {\em Parity–time symmetry and exceptional points in photonics}, \href{https://www.nature.com/articles/s41563-019-0304-9}{Nat. Mat. {\bf 18}, 783 (2019)}.

\bibitem{Fleury2015}
R. Fleury, D. Sounas, and A. Al\`{u}, {\em An invisible acoustic sensor based on parity-time symmetry}, \href{https://www.nature.com/articles/ncomms6905}{Nat. Commun. {\bf 6}, 5905 (2015)}. 

\bibitem{Helbig2020}
T. Helbig, T. Hofmann, S. Imhof, M. Abdelghany, T. Kiessling, L. W. Molenkamp, C. H. Lee, A. Szameit, M. Greiter, and R. Thomale, {\em Generalized bulk–boundary correspondence in non-Hermitian topolectrical circuits}, \href{https://www.nature.com/articles/s41567-020-0922-9}{Nat. Phys. {\bf 16}, 747 (2020)}.

\bibitem{Ghatak2020}
A. Ghatak, M. Brandenbourger, J. van Wezel, and C. Coulais, {\em Observation of non-Hermitian topology and its bulk–edge correspondence in an active mechanical metamaterial}, \href{https://www.pnas.org/content/117/47/29561}{Proc. Natl. Acad. Sci. U.S.A. {\bf 117}, 29561 (2020)}.

\bibitem{Kreibich2014}
M. Kreibich, J. Main, H. Cartarius, and G. Wunner, {\em Realizing PT-symmetric non-Hermiticity with ultracold atoms and Hermitian multiwell potentials}, \href{https://journals.aps.org/pra/abstract/10.1103/PhysRevA.90.033630}{Phys. Rev. A {\bf 90}, 033630 (2014)}.

\bibitem{koziifu}
V. Kozii and L. Fu, {\em Non-Hermitian Topological Theory of Finite-Lifetime Quasiparticles: Prediction of Bulk Fermi Arc Due to Exceptional Point}, \href{https://arxiv.org/abs/1708.05841}{arXiv:1708.05841}.

\bibitem{Yoshida2018}
T. Yoshida, R. Peters, and N. Kawakami, {\em Non-Hermitian perspective of the band structure in heavy-fermion systems}, \href{https://journals.aps.org/prb/abstract/10.1103/PhysRevB.98.035141}{Phys. Rev. B {\bf 98}, 035141 (2018)}.

\bibitem{Yoshida2020}
T. Yoshida, R. Peters, N. Kawakami, and Y. Hatsugai, {\em Exceptional band touching for strongly correlated systems in equilibrium}, \href{https://academic.oup.com/ptep/article/2020/12/12A109/5875995}{Prog. Theor. Exp. Phys. ptaa059 (2020)}.

\bibitem{Bergholtz2019}
E. J. Bergholtz and J. C. Budich, {\em Non-Hermitian Weyl physics in topological insulator ferromagnet junctions}, \href{https://journals.aps.org/prresearch/abstract/10.1103/PhysRevResearch.1.012003}{Phys. Rev. Research {\bf 1}, 012003(R) (2019)}.

\bibitem{brody14}
D.C. Brody, {\em Biorthogonal quantum mechanics}, \href{http://dx.doi.org/10.1088/1751-8113/47/3/035305}{Journal of Physics A: Math. Theor. {\bf 47} 034305 (2014)}.

\bibitem{BerryDeg}
M. Berry,  {\em Physics of Nonhermitian Degeneracies}, \href{http://dx.doi.org/10.1023/B:CJOP.0000044002.05657.04}{Czechoslovak Journal of Physics {\bf 54}, 1039 (2004)}.

\bibitem{carlstroembergholtz}
J. Carlstr\"{o}m and E.J. Bergholtz, {\em Exceptional links and twisted Fermi ribbons in non-Hermitian systems}, \href{http://dx.doi.org/10.1103/PhysRevA.98.042114}{Phys. Rev. A {\bf 98}, 042114 (2018)}.

\bibitem{molina}
R.A. Molina and J. Gonzalez, {\em Surface and 3D Quantum Hall Effects from Engineering of Exceptional Points in Nodal-Line Semimetals}, \href{http://dx.doi.org/10.1103/PhysRevLett.120.146601}{Phys. Rev. Lett. {\bf 120}, 146601 (2018)}.

\bibitem{disorderlinesribbons}
K. Moors, A.A. Zyuzin, A.Y. Zyuzin, R.P. Tiwari and T.L. Schmidt, {\em Disorder-driven exceptional lines and Fermi ribbons in tilted nodal-line semimetals}, \href{http://dx.doi.org/10.1103/PhysRevB.99.041116}{Phys. Rev. B {\bf 99}, 041116(R) (2019)}.

\bibitem{ourknots}
J. Carlstr{\"o}m, M. St\aa lhammar, J.C. Budich, and E.J. Bergholtz, {\em Knotted Non-Hermitian Metals}, \href{http://dx.doi.org/10.1103/PhysRevB.99.161115}{Phys. Rev. B {\bf 99}, 161115(R) (2019)}.

\bibitem{Ronnyknot1}
C.H. Lee, G. Li, Y. Liu, T. Tai, R. Thomale, and X. Zhuang, {\em Tidal surface states as fingerprints of non-Hermitian nodal knot metals}, \href{http://dx.doi.org/10.1038/s42005-021-00535-1}{Communications Physics {\bf 4}, 47 (2021)}.

\bibitem{ourknots2}
M. St\aa lhammar, L. R\o dland, G. Arone, J.C. Budich and E.J. Bergholtz, {\em Hyperbolic nodal band structures and knot invariants}, \href{http://dx.doi.org/10.21468/SciPostPhys.7.2.019}{SciPost Physics {\bf 7}, 019 (2019)}.

\bibitem{symprotnod}
J.C. Budich, J. Carlstr{\"o}m, F.K. Kunst, and E.J. Bergholtz, {\em Symmetry-protected nodal phases in non-Hermitian systems}, \href{http://dx.doi.org/10.1103/PhysRevB.99.041406}{Phys. Rev. B {\bf 99}, 041406(R) (2019)}.

\bibitem{expknots}
K. Wang, L. Xiao, J.C. Budich, W.Yi, and P. Xue, {\em Simulating Exceptional Non-Hermitian Metals with Single-Photon Interferometry}, \href{https://journals.aps.org/prl/abstract/10.1103/PhysRevLett.127.026404}{Phys. Rev. Lett. {\bf 127}, 026404 (2021)}.

\bibitem{topphot} 
L. Lu, J.D. Joannopoulos, and M. Solja\v ci\v c, {\em Topological Photonics}, \href{http://dx.doi.org/10.1038/nphoton.2014.248}{Nature Photonics {\bf 8} (2014)}.

\bibitem{speclat} 
B.A. Bell, K. Wang, A.S. Solntsev, D.N. Neshev, A.A. Sukhorukov, and B.J. Eggleton, {\em Spectral photonic lattices with complex long-range coupling}, \href{http://dx.doi.org/10.1364/OPTICA.4.001433}{Optica {\bf 4}, 11 (2017)}.

\bibitem{Ozawa2019}
T. Ozawa, H.M. Price, A. Amo, N. Goldman, M. Hafezi, L. Lu, M.C. Rechtsman, D. Schuster, J. Simon, O. Zilberberg, and I. Carusotto, {\em Topological photonics}, \href{https://journals.aps.org/rmp/abstract/10.1103/RevModPhys.91.015006}{Rev. Mod. Phys. {\bf 91}, 015006 (2019)}.

\bibitem{Bender1998} 
C.M. Bender and S. Boettcher, {\em Real Spectra in Non-Hermitian Hamiltonians Having $\mathcal{PT}$ Symmetry}, \href{https://journals.aps.org/prl/abstract/10.1103/PhysRevLett.80.5243}{Phys. Rev. Lett. {\bf 80}, 5243 (1998)}.

\bibitem{Volovik1999}
G. E. Volovik, {\em Simulation of a Panlev\'{e}-Gullstrand black hole in a thin $^3$He-A film}, \href{https://link.springer.com/article/10.1134/1.568079}{JETP Lett. {\bf 69}, 705 (1999)}.

\bibitem{PW2000}
M.K. Parikh and F. Wilczek, {\em Hawking Radiation as Tunneling}, \href{http://dx.doi.org/10.1103/PhysRevLett.85.5042}{Phys. Rev. Lett. {\bf 85}, 5042 (2000)}.

\bibitem{Schwarzschild}
K. Schwarzschild, {\em Uber das Gravitationsfeld eines Massenpunktes nach der Einsteinschen Theorie},  \href{http://old.phys.huji.ac.il/~barak_kol/Courses/Black-holes/reading-papers/SchwarzschildTranslated.pdf}{Sitzungsberichte der Koniglich Preussischen Akademie der Wissenschaften: Vol. {\bf 3}, pp. 189-196, (1916)}.

\bibitem{Painleve}
P. Painlevé, {\em La mécanique classique et la théorie de la relativité}, \href{https://gallica.bnf.fr/ark:/12148/bpt6k31267/f677.image}{C. R. Acad. Sci. (Paris) {\bf 173}, (1921)}.

\bibitem{Lemaitre}
G. Lema\^{i}tre, {\em L'Univers en expansion},  \href{https://ui.adsabs.harvard.edu/abs/1933ASSB...53...51L}{Annales de la Société Scientifique de Bruxelles, {\bf A53}, 51 (1933)}.

\bibitem{Vanzo2011}
L. Vanzo, G. Acquaviva, R. Di Criscienzo, {\em Tunnelling methods and Hawking's radiation: achievements and prospects},  \href{https://iopscience.iop.org/article/10.1088/0264-9381/28/18/183001}{Class. Quantum Grav., {\bf 28} 183001, (2011)}.

\bibitem{carroll}
S. M. Carroll, {\em Spacetime and Geometry: An Introduction to General Relativity}, Addison-Wesley (2003), ISBN-13 978-0805387322.

\bibitem{JennieTraschen}
J. Traschen, {\em An introduction to Black Hole evaporation}, Mathematical Methods of Physics, proceedings of the 1999 Londrina Winter School, editors A. Bytsenko and F. Williams, World Scientific (2000) \href{https://arxiv.org/abs/gr-qc/0010055}{arXiv:0010055}.

\bibitem{KrausKeski}
P. Kraus, E. Keski-Vakkuri, {\em Microcanonical D-branes and Back Reaction}, \href{http://dx.doi.org/10.1016/S0550-3213(97)00085-0}{Nucl. Phys. B {\bf 491}, 249 (1997)}.


\bibitem{Volovik2014} 
G.E. Volovik, and M.A. Zubkov {\em Emergent Weyl spinors in multi-fermion systems}, \href{http://dx.doi.org/10.1016/j.nuclphysb.2014.02.018}{Nuclear Physics, B {\bf 881} (2014)}.



\bibitem{Horova2005} 
P. Ho\v{r}ava, {\em Stability of Fermi Surfaces and 
K-Theory}, \href{http://dx.doi.org/10.1103/PhysRevLett.95.016405}{Phys. Rev. Lett. {\bf 95}, 016405 (2005)}.

\bibitem{Ortin} 
T. Ortín, {\em Gravity and Strings}, \href{https://doi.org/10.1017/CBO9780511616563}{Cambridge University Press (Cambridge Monographs on Mathematical Physics), (2004)}.

\bibitem{bfermiarcs} 
H. Zhou, C. Peng, Y. Yoon, C. W. Hsu, K. A. Nelson, L. Fu, J. D. Joannopoulos, M. Solja\v ci\v c, and B. Zhen, {\em Observation of bulk Fermi arc and polarization half charge from paired exceptional points}, \href{http://dx.doi.org/10.1126/science.aap9859}{Science {\bf 359}, 6379 (2018)}.

\bibitem{BLC1} 
D. Bernard and A. LeClair, "A classification of non-hermitian random matrices”, in {\em Statistical Field Theories}, edited by A. Cappelli and G. Mussardo (Springer Netherlands, Dordrecht, 2002) pp. 207-214.

\bibitem{BLC2}
D. Bernard and A. LeClair, {\em A Classification of Non-Hermitian Random Matrices}, \href{https://arxiv.org/abs/cond-mat/0110649}{arXiv:0110649}.

\bibitem{herring} 
C. Herring, {\em Accidental Degeneracy in the Energy Bands of Crystals}, \href{https://journals.aps.org/pr/abstract/10.1103/PhysRev.52.365}{Phys. Rev. {\bf 52} 365 (1937)}. 

\bibitem{branchCutSemimetals}
M. Ezawa, {\em Topological semimetals carrying arbitrary Hopf numbers: Fermi surface topologies of a Hopf link, Solomon's knot, trefoil knot, and other linked nodal varieties}, \href{https://journals.aps.org/prb/abstract/10.1103/PhysRevB.96.041202}{Phys. Rev. B {\bf 96}, 041202(R) (2017)}.

\bibitem{Nodallinksemimetals}
Z. Yan, R. Bi, H. Shen, L. Lu, S.-C. Zhang, and Z. Wang, \href{https://journals.aps.org/prb/abstract/10.1103/PhysRevB.96.041103}{\em Nodal-link semimetals}, {Phys. Rev. B {\bf 96}, 041103(R) (2017)}.

\bibitem{explink}
G. Chang, S.-Y. Xu, X. Zhou, S.-M. Huang, B. Singh, B. Wang, I. Belopolski, J. Yin, S. Zhang, A. Bansil, H. Lin, and M.Z. Hasan, {\em Topological Hopf and Chain Link Semimetal States and Their Application to Co$_2$MnGa}, \href{https://journals.aps.org/prl/abstract/10.1103/PhysRevLett.119.156401}{Phys. Rev. Lett. {\bf 119}, 156401 (2017)}.

\bibitem{floquetlinks}
L. Li, C.H. Lee, and J. Gong, {\em Realistic Floquet Semimetal with Exotic Topological Linkages
between Arbitrarily Many Nodal Loops}, \href{https://journals.aps.org/prl/abstract/10.1103/PhysRevLett.121.036401}{Phys. Rev. Lett. {\bf 121}, 036401 (2018)}.

\bibitem{nodalknotsemimetals}
R. Bi, Z. Yan, L. Lu and Z. Wang, {\em Nodal-knot semimetals}, \href{https://journals.aps.org/prb/abstract/10.1103/PhysRevB.96.201305}{Phys. Rev. B {\bf 96}, 201305 (2017)}.

\bibitem{HOEP}
P. Delplace, T. Yoshida, and Y. Hatsugai, {\em Symmetry-Protected Multifold Exceptional Points and Their Topological Characterization}, \href{https://journals.aps.org/prl/abstract/10.1103/PhysRevLett.127.186602}{Phys. Rev. Lett. {\bf 127}, 186602 (2021)}.

\bibitem{PageI}
Don N. Page, {\em Particle emission rates from a black hole: Massless particles from an uncharged, nonrotating hole}, \href{https://journals.aps.org/prd/abstract/10.1103/PhysRevD.13.198}{Phys. Rev. D {\bf 13}, 198  (1976)}.

\bibitem{PageII}
Don N. Page, {\em Particle emission rates from a black hole. II. Massless particles from a rotating hole}, \href{https://journals.aps.org/prd/abstract/10.1103/PhysRevD.14.3260}{Phys. Rev. D {\bf 14}, 3260  (1976)}.

\bibitem{massHR}
A. Coutant, A. Fabbri, R. Parentani, R. Balbinot, and P.R. Anderson, {\em Hawking radiation of massive modes and undulations}, \href{http://dx.doi.org/10.1103/PhysRevD.86.064022}{Phys. Rev. D {\bf 86} 064022 (2012)}.

\bibitem{NHPC}
F. Zhong, K. Ding, Y. Zhang, S. Zhu, C.T. Chan, and H. Liu, {\em Angle-Resolved Thermal Emission Spectroscopy Characterization of Non-Hermitian Metacrystals}, \href{https://journals.aps.org/prapplied/abstract/10.1103/PhysRevApplied.13.014071}{Phys. Rev. Applied {\bf 13}, 014071 (2020)}.

\bibitem{Chen2016}
W.-J. Chen, M. Xiao, and C.T. Chan, {\em Photonic crystals possessing multiple Weyl points and the experimental observation of robust surface states}, \href{https://www.nature.com/articles/ncomms13038#citeas}{Nat. Commun. {\bf 7}, 13038 (2016)}.

\bibitem{Yang2017}
B. Yang, Q. Guo, B. Tremain, L. E. Barr, W. Gao, H. Liu, B. B\'{e}ri, Y. Xiang, D. Fan, A. P. Hibbins, and S. Zhang, {\em Direct observation of topological surface-state arcs in photonic metamaterials}, \href{https://www.nature.com/articles/s41467-017-00134-1}{Nat. Commun. {\bf 8}, 97 (2017)}.

\bibitem{Lindblad1976}
G. Lindblad, {\em On the generators of quantum dynamical semigroups}, \href{https://link.springer.com/article/10.1007/BF01608499}{Commun. Math. Phys. {\bf 48}, 119 (1976)}.

\bibitem{Daley2014}
A. J. Daley, {\em Quantum trajectories and open many-body quantum systems}, \href{https://www.tandfonline.com/doi/abs/10.1080/00018732.2014.933502?journalCode=tadp20}{J. Adv. Phys. {\bf 63}, 77 (2014)}.

\bibitem{Bernier2017}
N. R. Bernier, L. D. T\'{o}th, A. Koottandavida, M. A. Ioannou, D. Malz, A. Nunnenkamp, A. K. Feofanov, and T. J. Kippenberg, {\em Nonreciprocal reconfigurable microwave optomechanical circuit}, \href{https://www.nature.com/articles/s41467-017-00447-1#citeas}{Nat. Commun. {\bf 8}, 604 (2017)}.

\bibitem{Malz2018}
D. Malz, L. D. T\'{o}th, N. R. Bernier, A. K. Feofanov, T. J. Kippenberg, and A. Nunnenkamp, {\em Quantum-Limited Directional Amplifiers with Optomechanics}, \href{https://journals.aps.org/prl/abstract/10.1103/PhysRevLett.120.023601}{Phys. Rev. Lett. {\bf 120}, 023601 (2018)}.

\bibitem{Metelmann2015}
A. Metelmann and A. A. Clerk, {\em Nonreciprocal Photon Transmission and Amplification via Reservoir Engineering}, \href{https://journals.aps.org/prx/abstract/10.1103/PhysRevX.5.021025}{Phys. Rev. X {\bf 5}, 021025 (2015)}.

\bibitem{Lin2003}
S.Y. Lin, J. Moreno, and J.G. Fleming, {\em Three-dimensional photonic-crystal emitter for thermal photovoltaic power generation}, \href{https://aip.scitation.org/doi/10.1063/1.1592614}{Appl. Phys. Lett. {\bf 83}, 380 (2003)}.

\bibitem{Florescu2005}
M. Florescu, H. Lee, A.J. Stimpson, and J. Dowling, {\em Thermal emission and absorption of radiation in finite inverted-opal photonic crystals}, \href{https://journals.aps.org/pra/abstract/10.1103/PhysRevA.72.033821}{Phys. Rev. A {\bf 72}, 033821  (2005)}.


\bibitem{Bekenstein}
J. D. Bekenstein, {\em Black Holes and Entropy}, \href{https://journals.aps.org/prd/abstract/10.1103/PhysRevD.7.2333}{Phys. Rev. D {\bf 7}, 2333 (1973)}.

\bibitem{Entanglemententropy}
S. N. Solodukhin, {\em Entanglement Entropy of Black Holes}, \href{https://link.springer.com/article/10.12942/lrr-2011-8}{Living Reviews in Relativity {\bf 14}, 8 (2011)}.



\bibitem{Feinberg1967}
G. Feinberg, {\em Possibility of Faster-Than-Light Particles}, \href{https://journals.aps.org/pr/abstract/10.1103/PhysRev.159.1089}{Phys. Rev. {\bf 159}, 1089 (1967)}.

\bibitem{Sticlet2021}
D. Sticlet, B. D\'{o}ra, C. P. Moca, {\em Kubo formula for non-Hermitian systems and tachyon optical conductivity}, \href{https://arxiv.org/abs/2104.02428}{arXiv:2104.02428}.

\bibitem{NHholography}
D. Arean, K. Landsteiner, and I. Salazar Landea, {\em Non-Hermitian holography}, \href{https://scipost.org/10.21468/SciPostPhys.9.3.032}{SciPost Phys. {\bf 9}, 032 (2020)}.

\bibitem{3EPemil}
I. Mandal, and E.J. Bergholtz, {\em Symmetry and Higher-Order Exceptional Points}, \href{https://journals.aps.org/prl/abstract/10.1103/PhysRevLett.127.186601}{Phys. Rev. Lett. {\bf 127}, 186601 (2021)}.

\bibitem{4EPMarcus}
M. St\aa lhammar, and E.J. Bergholtz, {\em Classification of exceptional nodal topologies protected by $\mathcal{PT}$ symmetry}, \href{https://journals.aps.org/prb/abstract/10.1103/PhysRevB.104.L201104}{Phys. Rev. B {\bf 104}, L201104 (2021)}.

\bibitem{Sundaram2021}
S. Sundaram, C.P. Burgess, and D.H.J. O'Dell, {\em Fall-to-the-centre as a $\mathcal{PT}$ symmetry breaking transition}, \href{https://arxiv.org/abs/2107.01511}{arXiv:2107.01511}.


\end{thebibliography}
\end{document}